\documentclass[a4paper,11pt]{article}
\pdfoutput=1
\usepackage{jcappub}
\usepackage{slashed}
\usepackage[section]{placeins}
\usepackage[utf8]{inputenc}
\newcommand{\Trh}{T_\text{RH}}
\newcommand{\Tmax}{T_\text{max}}
\newcommand{\gs}{g_\star}
\newcommand{\gss}{g_{\star s}}
\newcommand{\mpl}{M_\text{Pl}}

\begin{document}

\title{Non-minimally Coupled\\Vector Boson Dark Matter}
\author[a]{Basabendu Barman,}
\author[a]{Nicolás Bernal,}
\author[b]{\\Ashmita Das,}
\author[c,\,d]{and Rishav Roshan}

\affiliation[a]{Centro de Investigaciones, Universidad Antonio Nariño\\
Carrera 3 este \# 47A-15, Bogotá, Colombia}

\affiliation[b]{Centre for Strings, Gravitation and Cosmology, Department of Physics, Indian Institute of Technology Madras, Chennai 600036, India}

\affiliation[c]{Department of Physics, IIT Guwahati,  Guwahati 781039, India}

\affiliation[d]{Physical Research Laboratory, Ahmedabad - 380009, Gujarat, India}
\emailAdd{basabendu88barman@gmail.com}
\emailAdd{nicolas.bernal@uan.edu.co}
\emailAdd{ashmita.phy@gmail.com}
\emailAdd{rishav@prl.res.in}
\abstract{We consider a simple abelian vector dark matter (DM) model, where {\it only} the DM $(\widetilde{X}_\mu)$ couples non-minimally to the scalar curvature $(\widetilde{R})$ of the background spacetime via an operator of the form $\sim \widetilde{X}_\mu\,\widetilde{X}^\mu\,\widetilde{R}$. By considering the standard freeze-out scenario, we show, it is possible to probe such a non-minimally coupled DM in direct detection experiments for a coupling strength $\xi\sim\mathcal{O}\left(10^{30}\right)$ and DM mass $m_X\lesssim 55$~TeV, satisfying Planck observed relic abundance and perturbative unitarity. We also discuss DM production via freeze-in, governed by the non-minimal coupling, that requires $\xi\lesssim 10^{-5}$
to produce the observed DM abundance over a large range of DM mass depending on the choice of the reheating temperature. We further show, even in the absence of the non-minimal coupling, it is possible to produce the whole observed DM abundance via 2-to-2 scattering of the bath particles mediated by massless gravitons. 
}

\begin{flushright}
  PI/UAN-2021-697FT
\end{flushright}
\maketitle

\section{Introduction}
\label{sec:intro}
The existence of dark matter (DM) has been extensively proven from several astrophysical~\cite{Zwicky:1933gu, Zwicky:1937zza, Rubin:1970zza, Clowe:2006eq} and cosmological~\cite{Hu:2001bc, Aghanim:2018eyx} evidences (for a review, see, e.g. Refs.~\cite{Jungman:1995df, Bertone:2004pz, Feng:2010gw}). All these evidences unequivocally point towards the gravitational interactions of the DM. As far as its fundamental particle nature goes, it is already established from observations that DM has to be electrically neutral and stable at least at the scale of lifetime of the Universe. The measurement of the anisotropy in the cosmic microwave background (CMB) radiation provides the most precise measurement of the DM relic density, usually expressed as $\Omega_\text{DM}h^2 \simeq 0.12$~\cite{Aghanim:2018eyx}, which is an important constraint to abide by. Since the Standard Model (SM) of particle physics fails to offer a viable candidate, one has to look beyond the realms of the SM to explain the particle DM. The weakly interacting massive particle (WIMP)~\cite{Jungman:1995df, Kolb:1990vq} by far is the most popular DM candidate where one assumes the DM particles to be in thermal equilibrium in the early Universe due to strong enough coupling with the visible sector that gives rise to an interaction strength of the order of the weak scale. The DM abundance then freezes out once the interaction rate falls out of equilibrium as the Universe expands and cools down. The weak scale interaction strength with the visible sector provides a window for WIMP-like DM candidates to be probed in collider, indirect or scattering experiments (see, e.g., Ref.~\cite{Arcadi:2017kky}), however no significant excess over the background has been found so far to guarantee a potential discovery in either of these experiments. 

Contrary to the vanilla WIMP-paradigm, it is also possible that the DM particle couples to the visible sector very weakly, so that chemical equilibrium is never achieved. The DM is then produced by decay or annihilation processes from the visible sector, until the production ceases due to the cooling of the primordial thermal bath below the relevant mass scale connecting the DM particle to the visible sector. Due to the super weak coupling strength, the DM particles produced via the freeze-in mechanism are referred to as feebly interacting massive particles (FIMP)~\cite{McDonald:2001vt, Choi:2005vq, Kusenko:2006rh, Petraki:2007gq, Hall:2009bx, Bernal:2017kxu}. The feeble interaction strength between the DM and the SM sector in freeze-in scenario implies that these classes of models are inherently very difficult to search for in direct detection, indirect detection, or collider experiments. It has further been pointed out, depending on whether the DM interaction with the visible sector is renormalizable or non-renormalizable, freeze-in can be either infrared (IR) where the DM abundance becomes important at a low temperature~\cite{McDonald:2001vt, Hall:2009bx, Chu:2011be, Bernal:2017kxu, Duch:2017khv, Biswas:2018aib, Barman:2019lvm, Heeba:2018wtf}
or ultra-violate (UV) where the DM genesis takes place at the highest temperature achieved by the thermal bath~\cite{Hall:2009bx, Elahi:2014fsa, Chen:2017kvz, Bernal:2019mhf, Biswas:2019iqm, Barman:2020plp, Barman:2020ifq, Bernal:2020bfj, Bernal:2020qyu, Barman:2021tgt}.\footnote{This temperature can be the reheating temperature in the case of a sudden inflaton decay, but can also be much larger if the decay of the inflaton is non-instantaneous~\cite{Giudice:1999fb, Giudice:2000ex}.}   

Since all the confirmed evidences for DM simply suggest that DM should
have gravitational interaction, the production of gravitational DM and its detection prospects have been widely studied in the literature in the context of scalar, fermion and vector boson DM ~\cite{Chung:1998zb, Chung:2004nh, Lerner:2009xg, Ren:2014mta, Cata:2016dsg, Tang:2016vch, Cata:2016epa,  Kolb:2017jvz, Garny:2017kha, Ema:2018ucl, Hashiba:2018tbu, Ema:2019yrd, Barman:2021ugy, Ahmed:2020fhc, Bezrukov:2020wnl, Gross:2020zam}. This production mechanism of DM refers to the particle creation due to the time varying scale factor of the Universe \cite{Chung:1998zb,Parker:1969au, Birrell:1982ix, Chung:2001cb}. The production of ``supermassive" DM during the transition between an inflationary and a  matter-dominated (or radiation-dominated) Universe due to the (non-adiabatic) expansion of the background spacetime has been discussed in Refs.~\cite{Chung:1998zb, Chung:2001cb}. On a different note, Refs.~\cite{Garny:2015sjg, Tang:2016vch, Tang:2017hvq, Garny:2017kha, Mambrini:2021zpp, Barman:2021ugy} have studied the production of gravitational DM where only gravity {\it minimally} couples the DM to the visible sector, such as, via the annihilation of the SM bath particles and/or inflatons mediated via $s$-channel graviton exchange. 
Beyond the minimally coupled scenario, it is also possible that the DM is non-minimally coupled to gravity, characterised by the non-minimal coupling $\xi$, where $\xi=0$ mimics the minimally coupled scenario. In this context, it is worth to be mentioned that $\xi=1/6$ is known as  conformal coupling for a massless scalar field.
In case where the conformal symmetry is broken, $\xi$ can well be considered to be a free parameter\footnote{In case the SM Higgs is non-minimally coupled to gravity, $|\xi_h|\lesssim 2.6\times 10^{15}$~\cite{Atkins:2012yn, Xianyu:2013rya}.} that determines the DM-SM interaction strength, albeit suppressed by the Planck mass. While the phenomenology of non-minimally coupled DM has been studied both in the context of WIMP~\cite{Ren:2014mta, Sun:2020dla} and FIMP~\cite{Kaneta:2021pyx}, considering DM with or without any intrinsic spin, the study of non-minimally coupled vector DM is rather hard to find. 

In this work we have considered a simple scenario, where a massive vector boson DM $(X_\mu)$ that originates from an abelian gauge extension of the SM, has a non-minimal coupling to the gravity. Here we consider {\it only} the DM to be non-minimally coupled to gravity which makes our model construction very economical in terms of the number of free parameters. Considering dimension-4 operators of the form $\sim \widetilde{X}_\mu \widetilde{X}^\mu \widetilde{R}$, we show the vector DM can either undergo pair-annihilation to the SM final states to produce the Planck observed relic abundance via freeze-out or can be produced from the scattering of the bath particles giving rise to out-of-equilibrium production via freeze-in. 
It is worth mentioning that a simple construction like $\sim \widetilde{X}_\mu \widetilde{X}^\mu \widetilde{R}$ is popular in the context of inflation~\cite{Golovnev:2008cf, Golovnev:2008hv, Maleknejad:2011sq, Bertolami:2015wir, Oliveros:2016myr}, whereas its prominence is rarely explored in the context of DM Physics.
Considering the vector DM to be a standard WIMP, for the freeze-out scenario, one can have a viable parameter space safe from the stringent (spin-independent) direct search bounds for $\xi \sim \mathcal{O}(10^{30})$. We emphasise that in previous works on non-minimally coupled scalar DM, the DM communicates with the SM via a (non-)standard Higgs mediator (which is also non-minimally coupled), that suppresses its direct search cross-section due to heavy mediator mass or small momentum exchange~\cite{Ren:2014mta,Sun:2020dla}. This is in sharp contrast to the present scenario where the absence of mediator opens up direct search possibilities via contact interactions. The presence of the non-minimally coupled scalar mediator, on the other hand, can also provide observable signatures for DM indirect detection in terms of gamma-ray flux, anti-proton flux or positron excess that in turn further constrain the DM mass and the non-minimal coupling. The absence of such mediators set the present model free from those bounds. Additionally, the freeze-in production can occur both in the presence or absence of the non-minimal coupling. In contrast to the freeze-out case, freeze-in production of DM via non-minimal coupling requires $\xi$ to be smaller by several orders of magnitude, that ensures the DM production rate remains below the Hubble expansion parameter. For typical choices of the reheating temperature and the non-minimal coupling $\xi$, it is possible to produce DM via freeze-in over a wide mass range starting from a few keV to several orders of TeV, satisfying various theoretical and observational bounds. Finally, we show, gravitational UV freeze-in, corresponding to the minimally coupled scenario, can lead to DM overabundance for a large reheating temperature. We thus consider DM genesis in the early Universe through: $i)$ freeze-out via non-minimal coupling, $ii)$ freeze-in via non-minimal coupling, and $iii)$ minimally coupled gravitational UV freeze-in, and in each case we illustrate the viable parameter space.


This paper is organised as follows. The model construct is discussed in Sec.~\ref{sec:framework} elaborating the underlying action in Jordan and in the Einstein frame. The WIMP phenomenology is discussed in Sec.~\ref{sec:u1param}, where the viable parameter space for the DM satisfying the bounds from Planck observed relic density and spin-independent direct detection is shown. We then move on to the discussion regarding freeze-in production of vector DM in Sec.~\ref{sec:fiu1x}, under which freeze-in via non-minimal coupling is addressed in subsection~\ref{sec:nonm-fi} and in subsection~\ref{sec:min-fi} the gravity mediated (minimal) DM production is addressed. Finally, we conclude in Sec.~\ref{sec:concl}. Appendices are provided for the paper to be self-sufficient.

\section{The framework}
\label{sec:framework}

\subsection{Non-minimally coupled vector DM}
\label{sec:model}
We consider the vector DM $X_\mu$ to be a massive gauge boson under some abelian $U\left(1\right)_X$ symmetry and construct the following action in the Jordan frame, where the DM is explicitly coupled to the scalar curvature of the background spacetime ($\widetilde{R}$)
\begin{equation} \label{eq:abjordan}
    \mathcal{\widetilde{S}} = \int d^4x\,\sqrt{-\widetilde{g}}\,\bigg[\frac{1}{2}\bigg(\mpl^2-\xi\,\widetilde{X}_{\mu}\widetilde{X}^{\mu}\bigg)\,\widetilde{R}+\,\mathcal{\widetilde{L}}_\text{DM}+\,\mathcal{\widetilde{L}}_\text{SM}\bigg],
\end{equation}
with
\begin{equation} \label{eq:u1ldm}
    \mathcal{\widetilde{L}}_\text{DM} = -\frac{1}{4}\widetilde{X}_{\mu\nu} \widetilde{X}^{\mu\nu} + \frac{1}{2}m_X^2 \widetilde{X}_\mu \widetilde{X}^\mu.
\end{equation}
Note that $\widetilde{R}$,  $\mathcal{\widetilde{L}}_\text{DM}$ and $\mathcal{\widetilde{L}}_\text{SM}$ (defined in Appendix~\ref{sec:jor-ein}) belong to the Jordan frame and defined with respect to the metric in the Jordan frame {\it i.e.}  $\widetilde{g}_{\mu\nu}$. $\mpl$ is the reduced Planck scale $\approx 2.4 \times 10^{18}$~GeV.
Here we assume a Stueckleberg mass term\footnote{In abelian gauge theories, the Stueckelberg mechanism can be taken as the limit of the Higgs mechanism where the mass of the real scalar is sent to infinity and only the pseudoscalar is present~\cite{Stueckelberg:1938hvi, Ruegg:2003ps, Kors:2004dx, Kors:2005uz}.} for the vector DM which prohibits any connection of the DM with the visible sector apart from gravity. This leads to the simplest scenario for a non-minimally coupled abelian vector DM.%
\footnote{ It is well known that the kinetic term of the longitudinal mode of a non-minimally coupled massive vector boson becomes negative on sub-horizon scale during inflation~\cite{Dvali:2007ks, Dimopoulos:2008yv, Karciauskas:2010as}. Such modes, called ghosts, are dangerous as they lead to vacuum decay, making these theories phenomenologically viable only as effective theories below certain cut-off scale~\cite{Himmetoglu:2008zp, Himmetoglu:2009qi}.  Several prescriptions have been proposed to cure this problem e.g., in Refs.~\cite{Karciauskas:2010as, Nakayama:2019rhg, Nakayama:2020rka} although no definite conclusion regarding the viability of such theories have been reached.}
We also assume the presence of an unbroken $\mathbb{Z}_2$ symmetry under which the DM is odd while all the SM fields are even, thus ensuring the stability of the DM by forbidding the kinetic mixing term. In the absence of the $\mathbb{Z}_2$ symmetry, the dark gauge boson can still account for all of the DM relic abundance if the kinetic mixing is of the order $\epsilon \lesssim \mathcal{O}\left(10^{-8}\right)$ for DM masses below twice the electron mass~\cite{Arias:2012az, Caputo:2021eaa}, else the cosmological stability condition requires even smaller values $(\lesssim 10^{-15})$ of the kinetic mixing parameter~\cite{Bloch:2016sjj, Lin:2019uvt}. 

Now, to obtain the form of the action in the Einstein frame, we perform a conformal transformation $g_{\mu\nu} = \omega^2\, \widetilde{g}_{\mu\nu}$ to the action $\mathcal{\widetilde{S}}$, where $\omega^2 = 1-\frac{\xi}{\mpl^2}\widetilde{X}_\mu \widetilde{X}^\mu$ and $g_{\mu\nu}$ stands for the spacetime metric in the Einstein frame.
This leads us to the action in the Einstein frame where the gravitational part of the action turns into the well-known Einstein-Hilbert action.
Thus, considering the metric signature $(+,-)$, and using the above conformal transformation, we obtain (see Appendix~\ref{sec:jor-ein} for details), 
\begin{eqnarray}
    \mathcal{S}= \int d^4x\,\sqrt{-g}&&\Biggl[\frac{\mpl^2\,R}{2} + \frac{3\,\omega^4}{4\,\mpl^{2}}\,\nabla_{\alpha}(\xi\,X_{\mu}X^{\mu})\,\nabla^{\alpha}(\xi\,X_{\mu}X^{\mu}) - \frac{1}{4}\,X_{\mu\nu}\,X^{\mu\nu}\nonumber\\
    && + \frac{1}{2\omega^2}m_X^2\,X_\mu\,X^\mu +\,\frac{1}{\omega^4}\,(\mathcal{L}_{Y}-V(H))+\,\frac{1}{\omega^2}(D_{\mu} H)^{\dagger}(D^{\mu} H)\nonumber\\
    && +\,\frac{i}{\omega^3}\,\bar{f}\,\gamma^a\,\partial_a\,f -\,\frac{1}{4}\,g^{\mu\alpha}\,g^{\nu\beta}F_{\mu\nu}^{(a)}\,F_{\alpha\beta}^{(a)}+\,\frac{3\,i}{\omega^4}\,\bar{f}\,(\slashed{\partial}\,\omega)\,f\Biggr],
\label{eq:abein2}
\end{eqnarray}
where $\mathcal{S}$ represents the action in the Einstein frame. Note that, all the parameters in the above action such as $R$, $\mathcal{L}_{\text{SM}}$ and as well as the DM sector, now belong to the Einstein frame and defined with respect to the metric $g_{\mu\nu}$. Now, expanding $\omega$ in the small field limit $\xi \left(X_\mu\,X^\mu\right) /\mpl^2 \ll 1$~\cite{Burgess:2010zq} we obtain
\begin{equation}
    \omega = \bigg(1+\frac{\xi}{\mpl^2}X_\mu X^\mu\bigg)^{-1/2}\simeq 1-\frac{\xi}{2\mpl^2}X_\mu X^\mu+\,\mathcal{O}\left(\mpl^{-4}\right).
\label{eq:omega}
\end{equation}
Using Eq.~(\ref{eq:omega}) we find
\begin{eqnarray}
    \mathcal{S}&=& \int d^4x \sqrt{-g}\Biggl[\underbrace{\frac{\mpl^2\,R}{2}}_{{\rm Pure\,gravity}}+\,\underbrace{(D_{\mu} H)^{\dagger}(D^{\mu} H)+\,(\mathcal{L}_{Y}-V(H))+\,i\,\bar{f}\,\gamma^\alpha\,\partial_\alpha\,f
    -\,\frac{1}{4}\,F_{\mu\nu}^{(a)}\,F^{(a)\mu\nu}}_{{\rm SM}}\nonumber\\
    &&\underbrace{-\,\frac{1}{4}\,X_{\mu\nu}\,X^{\mu\nu}+\frac{1}{2}m_X^2\,X_\mu\,X^\mu}_{{\rm Free\,DM\, sector}}+\,\bigg\{\frac{\xi\,m_{X}^{2}}{2\,\mpl^2}\,(X_{\alpha}X^{\alpha})^2+\,\frac{2\xi}{\mpl^2}\,X_{\alpha}X^{\alpha}\,(\mathcal{L}_{Y}-V(H))\nonumber\\
    &&+\,\frac{\xi}{\mpl^2}X_{\alpha}X^{\alpha}\,(D_{\mu} H)^{\dagger}(D^{\mu} H)+\,\frac{3 i\,\xi}{2\, \mpl^2}\,X_{\alpha}X^{\alpha}\,\bar{f}\,\gamma^\mu\,\partial_\mu\,f
    -\,\frac{3\,i\xi}{2 \mpl^2}\,\bar{f}\,\slashed{\partial}\,(X_{\alpha}X^{\alpha})\,f\nonumber\\
    &&+\,\frac{3\,\xi^2}{4\,\mpl^{2}}\,\nabla_{\alpha}(X_{\mu}X^{\mu})\,\nabla^{\alpha}(X_{\mu}X^{\mu})\bigg\}\biggr].
    \label{eq:abein2_1}
\end{eqnarray}
From the previous expression, it can be clearly seen that the terms within the curly bracket are  associated with $\xi$, and exhibit all possible DM-SM interactions in the theory. This also shows that it is not possible for the DM to decay gravitationally, contrary to Refs.~\cite{Cata:2016dsg, Cata:2016epa, Bezrukov:2020wnl}. Note that the mass term for the DM is also modified by the conformal transformation parameter. However, such a mass correction is negligible compared to the Stueckleberg mass because of Planck suppression.
Our choice of working with the Einstein frame (where a field minimally couples to gravity) relies on the fact that in the Jordan frame (where a field non-minimally couples to gravity), due to the non-canonicality of the gravitational Lagrangian, some physical parameters such as the stress energy tensor of the non-minimally coupled field turns out to be more complicated than in the Einstein frame~\cite{Faraoni:1998qx, Faraoni:1999hp}.
Furthermore, due to non-canonicality of the kinetic term of the gravitational field in the Jordan frame, one obtains the propagator for the graviton modified by the factor $\propto(\xi \tilde{X}^2)^{-1}$,  which appears in a more manageable form in the Einstein frame. Therefore, we consider the Einstein frame to be the physical frame, where the theoretical and observational predictions become comprehensible.

\subsection{Dark matter production mechanisms}
\label{sec:dm-prod}
It is clear from Eq.~\eqref{eq:abein2_1} that the interaction of the DM with the visible sector is determined by the non-minimal coupling strength $\xi$. Thus, depending on the strength, the DM in the present model can be produced via: $i)$ freeze-out, where the DM $X$ acts as a thermal WIMP that gives rise to the observed relic abundance for a  $\xi\sim\mathcal{O}\left(10^{30}\right)$ via 2-to-2 annihilation with the SM particles in the final states. Such large coupling is required to ensure that the DM remains in equilibrium with the thermal bath in the early Universe, and $ii)$ freeze-in, where the DM is produced from 2-to-2 scattering of the bath particles. Contrary to the freeze-out scenario, freeze-in requires $\xi\lesssim\mathcal{O}\left(10^5\right)$ to ensure the DM production takes place out of equilibrium at large temperatures. In the former case the DM can be as heavy as $m_X\lesssim 55$~TeV satisfying unitarity bound, while in the later case the vector DM becomes a typical FIMP and can be as heavy as $m_X\sim\mathcal{O}\left(10^{10}\right)$~GeV depending on the choice of the reheating temperature. Apart from these options, which depend on the non-minimal coupling strength, the DM can also be produced via the irreducible gravitational UV freeze-in, as gravitons can still mediate between the DM and the SM. Such a process is always present since this corresponds to the minimally coupled scenario with $\xi=0$. Tab.~\ref{tab:dm-prod} provides a summary of the different DM genesis mechanisms  addressed in this paper. 

\begin{table}[htb!]
\centering
\begin{tabular}{|l|c|c|c|c|c|c|c|c|c|r|}
\hline
DM production & $\xi$ \\
\hline
Freeze-out & $\sim\mathcal{O}\left(10^{30}\right)$ \\
(non-minimal) freeze-in & $\lesssim\mathcal{O}\left(10^5\right)$ \\
Gravitational UV freeze-in (minimal) & 0 \\
\hline
\end{tabular}
\caption{Typical non-minimal couplings and their corresponding DM production mechanisms.}
\label{tab:dm-prod}
\end{table}

The Boltzmann equation (BEQ) governing the DM number density $n_X$ evolution can be expressed as~\cite{Kolb:1990vq}
\begin{equation}
    \dot{n}_X+3\,H\,n_X=\gamma
\label{eq:BEQ}
\end{equation}
where $\gamma$ is the reaction rate density that depends on the underlying DM production mechanism, while the Hubble parameter reads $H(T) = \frac{\pi}{3}\sqrt{\frac{g_\star}{10}}\frac{T^2}{\mpl}$, in a SM radiation dominated Universe, and $\gs(T)$ corresponds to the number of relativistic degrees of freedom contributing to the SM radiation~\cite{Drees:2015exa}. It is convenient to express Eq.~\eqref{eq:BEQ} in terms of the dimensionless quantity $x=m_X/T$ 
\begin{equation}
    x\,H\,s\,\frac{dY_X}{dx}=\gamma(x)\,,
\label{eq:BEQ2}
\end{equation}
where we define the DM yield $Y_X\equiv n_X/s$, as the ratio of DM number density $n_X$ and the comoving entropy density in the visible sector is defined via $s(T) = \frac{2\pi^2}{45}g_{\star s} T^3$, with $\gss(T)$ being the number of relativistic degrees of freedom contributing to the SM entropy~\cite{Drees:2015exa}. To match the observed DM abundance $\Omega_\text{DM} h^2 \simeq 0.12$ at the present epoch $T=T_0$, the DM yield has to be fixed so that $m_X Y_X\left(T_0\right) = \Omega_\text{DM} h^2 \frac{1}{s_0} \frac{\rho_c}{h^2} \simeq 4.3 \times 10^{-10}$~GeV, where $\rho_c \simeq 1.1 \times 10^{-5} h^2$~GeV/cm$^3$ is the critical energy density and $s_0 \simeq 2.9 \times 10^3$~cm$^{-3}$ is the entropy density at present~\cite{Aghanim:2018eyx}.

\section{Freeze-out of vector dark matter}\label{sec:u1param}
The phenomenology of abelian vector boson DM has been widely studied in the literature both in the context of freeze-out and freeze-in production (see, e.g., Refs.~\cite{Servant:2002aq, Birkedal:2006fz, Farzan:2012hh, Baek:2012se, Bian:2013wna, Choi:2013qra, Baek:2013dwa, Baek:2013fsa, Baek:2014goa, Ko:2014gha, Baek:2014poa, Ko:2014loa, Duch:2015jta, Beniwal:2015sdl, Kamon:2017yfx, Duch:2017nbe, Duch:2017khv, Arcadi:2017jqd, Baek:2018aru, YaserAyazi:2019caf, Barman:2020ifq}). In all these cases the DM is minimally coupled to gravity and communicates with the visible sector via the Higgs portal.
Here, however, we consider that both sectors communicate only through gravity by considering a Stueckelberg mass term for the vector DM. In such a scenario,we end up with only two free parameters
\begin{equation}
    \{m_X,\, \xi\}
\end{equation}
that control the resulting parameter space for the DM. This makes the present model simple yet testable. 

\begin{figure}[htb!]
$$
\includegraphics[scale=0.35]{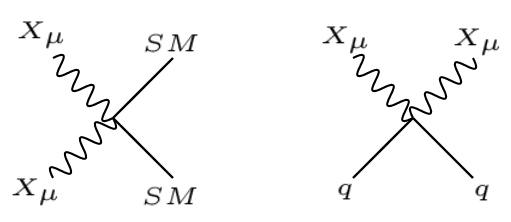}
$$
\caption{Left: Pair annihilation of abelian vector DM into SM final states leading to DM relic abundance, where SM stands for all SM particles. Right: Scattering of DM against SM quarks $q$ leading to spin-independent direct search.}\label{fig:ann-u1}
\end{figure}

The freeze-out parameter space for the DM is primarily constrained by the requirement of obtaining the observed relic abundance. One should note here, due to the absence of Higgs portal, the DM can annihilate to the SM final states only via contact interaction involving a pair of DM and a pair of SM particles as shown in Fig.~\ref{fig:ann-u1}. This contact interaction is induced solely by the gravity, and hence proportional to the strength of the non-minimal coupling $\xi$. The DM abundance is obtained by numerically solving the BEQ in Eq.~\eqref{eq:BEQ}, where
\begin{equation}
    \gamma = -\langle\sigma v\rangle\Bigl(n_X^2-n_{X\text{eq}}^2\Bigr)
\end{equation}
for a standard WIMP scenario~\cite{Edsjo:1997bg}, where $n_{X\text{eq}}$ is the equilibrium DM number density given by $n_{X\text{eq}}(T) \simeq \frac{T}{2\pi^2} g_X\, m_X^2\, K_2\left(\frac{m_X}{T}\right)$, for non-relativistic DM. In the present case, the pair annihilation cross-section for the DM to the SM final states (left panel of Fig.~\ref{fig:ann-u1}) is $s$-wave dominated:
\begin{equation}
\left(\sigma v\right)_{XX\to\text{SM}\text{SM}}\simeq \begin{cases}
        \frac{4N_c m_X^2\xi^2}{ \mpl^4\,\pi} \sqrt{1-x^2}\,\Bigl(4-x^2-3x^4\Bigr)+\,\mathcal{O}[v^2]& x\equiv m_f/m_X\\[8pt]
        \frac{\delta_V\,\xi^2}{144\,\pi m_X^2}\Bigl(\frac{g_2 v_d}{c_w \mpl}\Bigr)^4\sqrt{1-x^2}\,\Bigl(1+2x^2+3x^4\Bigr)+\,\mathcal{O}[v^2] &  x\equiv m_V/m_X\\[8pt]
       \frac{\xi^2 m_X^2}{9\,\pi\,\mpl^4}\sqrt{1-x^2}\,\Bigl(1+2x^4\Bigr)+\,\mathcal{O}[v^2]&
        x\equiv m_h/m_X
    \end{cases}\label{eq:ann-u1}
\end{equation}
where $N_c = 1\, (3)$ for leptonic (quark) final states, $g_2$ is the SM $SU(2)_L$ gauge coupling with $s_w$ as the sine of the weak mixing angle and $\delta_V = 1\, (2)$ for $Z$ $(W^\pm)$ boson final states. Here $v_h\simeq 246~\rm GeV$ is the Higgs vacuum expectation value and $v$ is the relative velocity between two incoming DM particles. The final DM abundance can be obtained by solving the BEQ numerically. However, for DM annihilations dominated by $s$-wave processes, the relic abundance can be approximated as~\cite{Kolb:1990vq}
\begin{equation}
    \Omega_X h^2 \simeq 1.07\times 10^9 \frac{x_f~\text{GeV}^{-1}}{\left(g_{\star s}/\sqrt{g_\star}\right)\,\langle\sigma v\rangle\,\mpl}\label{eq:omegaX}
\end{equation}
where $x_f=m_X/T_f$ is the freeze-out temperature that can be determined by the condition $H\left(x_f\right)=\langle\sigma v\rangle\left(x_f\right)$.

\begin{figure}[htb!]
$$
\includegraphics[scale=0.36]{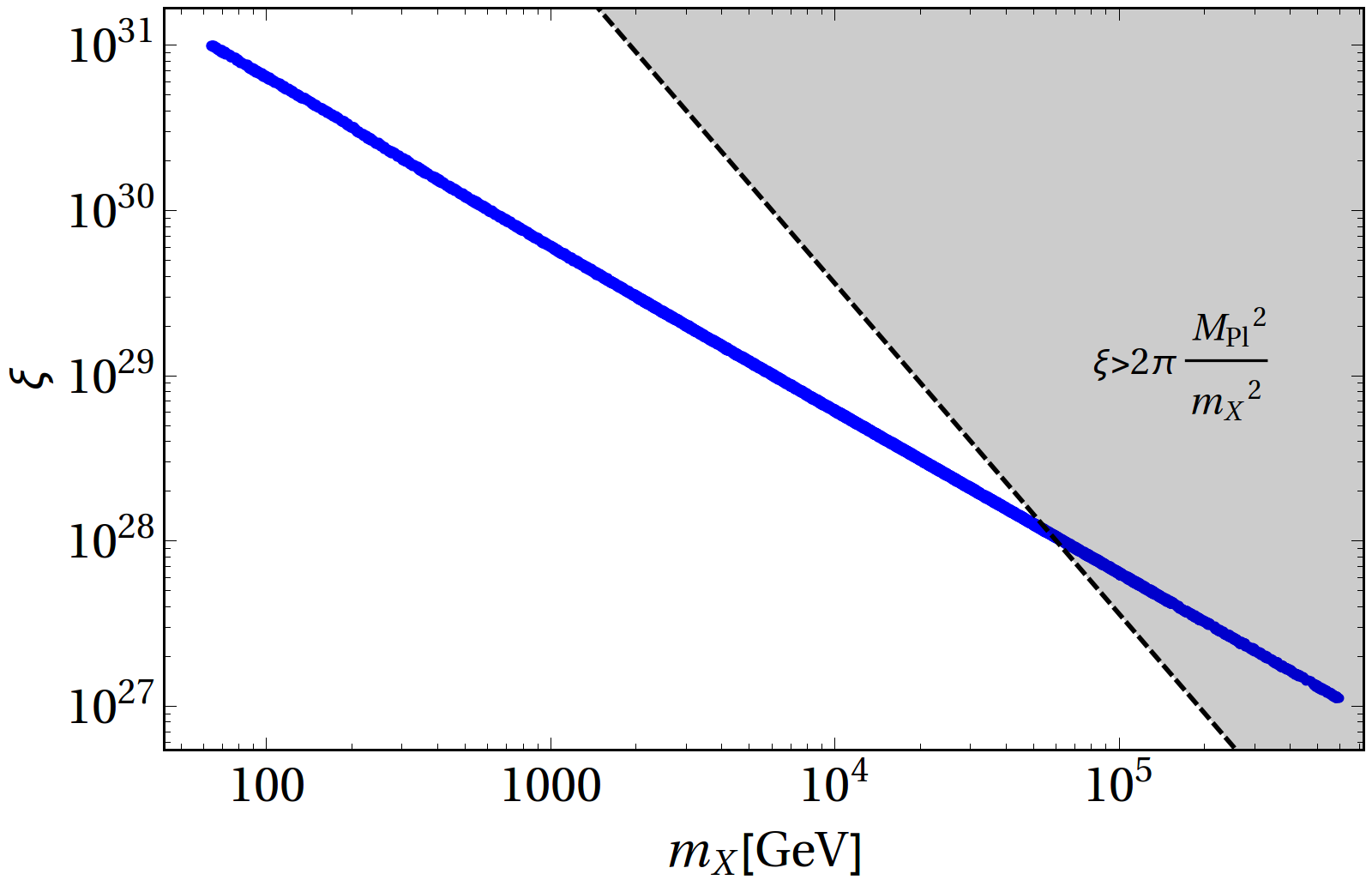}~~~~
\includegraphics[scale=0.36]{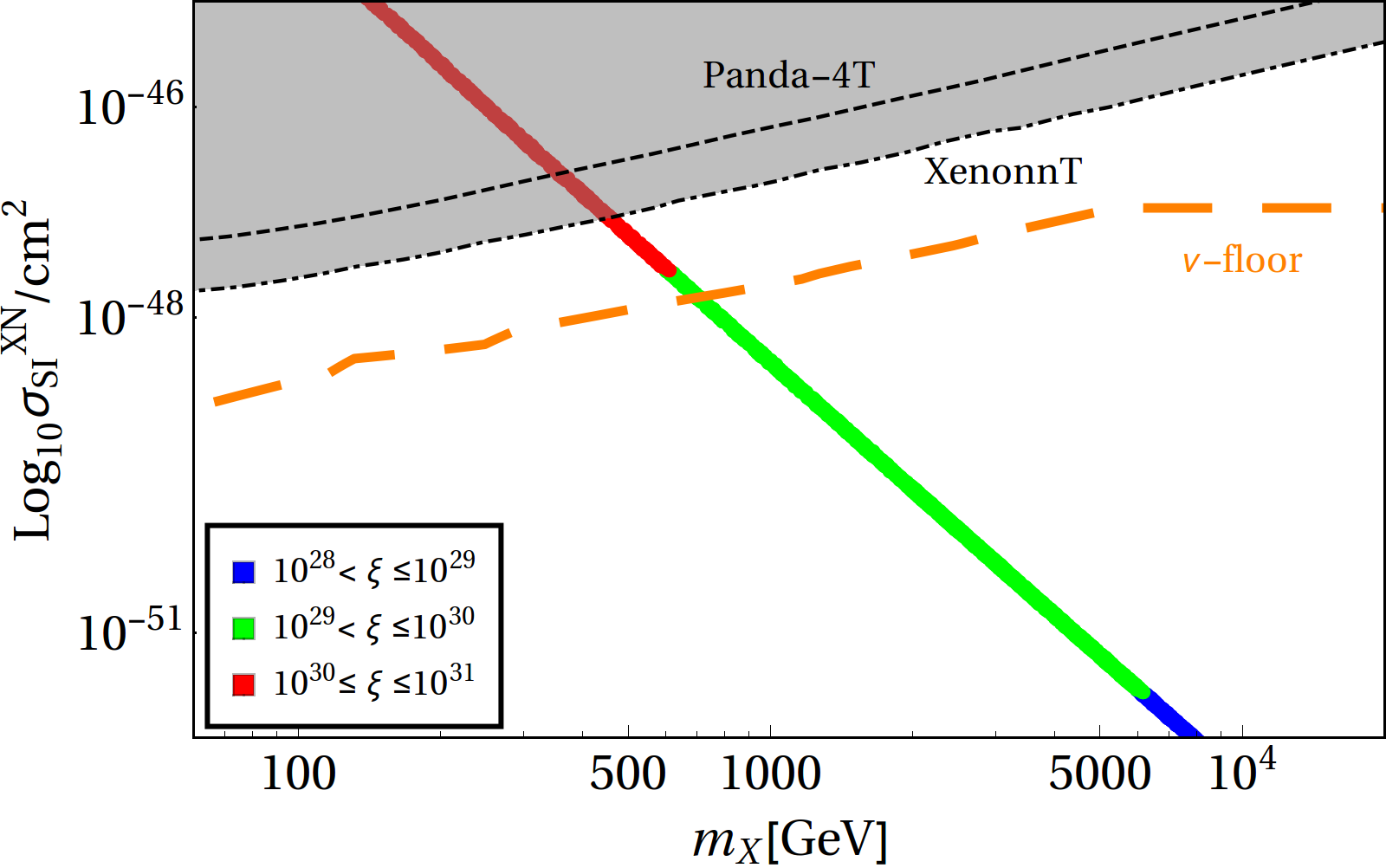}
$$
\caption{Left: The relic density allowed region is shown in the bi-dimensional $\xi-m_X$ plane. The grey shaded region to the right of the black dashed line is ruled out by the unitarity bound on WIMP mass. Right: The relic density satisfied points are shown in the direct search plane where different coloured regions correspond to different ranges of $\xi$. The spin-independent direct search limits from Panda-4T and projected XENONnT experiments are shown via black dashed and black dot-dashed curves respectively. The dashed orange curve indicates the expected discovery limit corresponding to the so called ``$\nu$-floor" from CE$\nu$NS of solar and atmospheric neutrinos for a Ge target.
}\label{fig:u1param}
\end{figure}

To obtain the freeze-out parameter space, we implemented this model in {\tt LanHEP}~\cite{Semenov:2008jy} and computed the relic abundance numerically in {\tt MicrOmegas}~\cite{Belanger:2010pz}. The values of $\xi$ required to match the whole observed DM abundance are shown in the left panel of Fig.~\ref{fig:u1param}. From Eq.~\eqref{eq:ann-u1} it is clear that the annihilation cross-section in all cases is proportional to $\xi^2/\mpl^4$. Thus, with an increase in the strength of the non-minimal coupling, one should expect DM under abundance, while reducing the non-minimal coupling should lead to over abundance.

The DM gives rise to a spin-independent direct search cross-section as shown in the right panel of Fig.~\ref{fig:ann-u1}. Even in the absence of any Higgs portal, the DM can still recoil against the SM quarks (via contact interaction), giving rise to direct detection signals that depend only on the non-minimal coupling $\xi$.
On top of relic abundance, the spin-independent direct search exclusion limit also puts very stringent bound on the DM parameter space, particularly constraining large values for the non-minimal coupling $\xi$. The effective DM-quark coupling can be parameterized as
\begin{equation}
    f_q \simeq \frac{\xi\,m_q}{\mpl^2}\,, 
\end{equation}
which gives rise to a DM-nucleus scattering cross-section for the vector boson DM~\cite{Hisano:2010yh, Hisano:2015bma}
\begin{equation}
    \sigma_\text{SI}^{XN} = \frac{\mu_{XN}^2}{\pi\,m_X^2} \left[Zf_p + \left(A - Z\right) f_n \right]^2\label{eq:dd-cs}
\end{equation}
where $f_{p,n}$ are the effective DM-nucleon coupling  and $\mu_{XN}=m_X\,m_N/\left(m_X+m_N\right)$ is the DM-nucleus reduced mass, with $Z$ being the number of protons and $A-Z$ the number of neutrons. We rely on the hadronic matrix elements and DM form factors included in {\tt MicrOmegas} to compute the direct search cross-section.
The right panel of Fig.~\ref{fig:u1param} shows the parameter space in the $\xi-m_X$ plane where Planck observed relic abundance is satisfied, together with constraints from spin-independent direct detection. Here we see, for a comparatively lower DM mass one needs a larger $\xi$ to satisfy the observed relic abundance. This can again be understood from Eq.~\eqref{eq:ann-u1}, where we find the total annihilation cross-section to SM final states goes roughly as $\left(\sigma v\right)_\text{total} \sim \mathcal{A}\, \frac{\xi^2\,m_X^2}{\mpl^4} \left(1 + \mathcal{C}\, \frac{\mpl^4}{m_X^4}\right)$ in the limit $x\to 0$, where $\mathcal{A} = \left(\frac{16N_c}{\pi} + \frac{1}{9\pi}\right)$ and $\mathcal{C} = \mathcal{B}/\mathcal{A}$ with $\mathcal{B} = \frac{\delta_v}{144\pi} \left(\frac{g_2\, v_d}{c_w\, \mpl}\right)^4$.
The relic abundance thus becomes $\Omega_X h^2\sim\frac{1}{\left(\sigma v\right)_\text{total}}\sim m_X^2\,\mpl^4/\left(\mathcal{A}\,\xi^2\left[m_X^4+\mathcal{B}\,\mpl^4\right]\right)$, which behaves as $\Omega_X h^2\sim 1/\left(\xi^2\,m_X^2\right)$ for $m_X\gtrsim 20~\rm GeV$. Therefore, a larger DM mass requires a smaller $\xi$ and vice versa, in order to obtain the observed relic abundance. This, in turn, influences the direct search allowed parameter space for the DM. This is seen from the right panel of Fig.~\ref{fig:u1param}, where higher values of $10^{30} \lesssim \xi \lesssim 10^{31}$ (red points) are discarded from the present limit from PandaX-4T experiment~\cite{PandaX-4T:2021bab} (black dashed curve) and mostly from future projection of XENONnT~\cite{XENON:2020kmp} up to DM mass of $m_X\sim 500~\rm GeV$.
For DM mass $m_X \gtrsim 500$~GeV the direct search bounds are relaxed (green points) since a smaller $\xi$ is needed to satisfy the desired abundance for larger DM mass as argued above, which in turn produces a smaller $\sigma_\text{SI}^{XN}$ aiding the direct search.
A large part of the viable parameter space, however, lies beyond the so called ``$\nu$-floor"~\cite{Billard:2013qya}, below which the number of neutrino events due to coherent elastic neutrino-nucleus scattering (CE$\nu$NS) is expected to be much larger than the number of DM events, which prevents to identify DM signals with certainty.  

Another constraint on the DM mass and non-minimal coupling strength can be derived from the requirement of perturbative unitarity. Here we restrict ourselves to the tree-level unitarity bound~\cite{Peskin:1995ev}
\begin{equation}
    \left|\text{Re}\left(a_J\right)\right|<\frac{1}{2} 
\end{equation}
where $a_J$ is the is the partial-wave amplitude for the total angular momentum $J$, and is related to the tree-level scattering amplitude $\mathcal{M}$ via
\begin{equation}
    a_J\left(s\right)=\frac{1}{32\pi}\int_{-1}^{+1}\,d\left(\cos\theta\right)\,P_J\left(\cos\theta\right)\,\mathcal{M}
\end{equation}
where $P_J\left(\cos\theta\right)$ is the Legendre polynomial of degree $J$. Here we provide an approximate analytical bound on the DM parameter space for both the freeze-out and freeze-in scenario. We note, the scattering amplitude in either case has a dependence of the from
\begin{equation}
    \mathcal{M} \simeq \xi\,\frac{s}{\mpl^2}\,.
\end{equation}
This leads to  
\begin{equation} \label{eq:unitarity}
    \sqrt{s} < \sqrt{\frac{8\pi}{\xi}}\,\mpl
\end{equation}
from the requirement of partial-wave unitarity of the $S$-matrix, which in turn constraints the annihilation cross section in the early Universe. A part of the parameter space for the WIMP-like DM is thus excluded, as shown by the grey shaded region in the left panel of Fig.~\ref{fig:u1param}. For $\xi\sim\mathcal{O}(10^{30})$, Eq.~\eqref{eq:unitarity} implies, the cut-off scale $\Lambda\sim\mathcal{M_\text{pl}}/\sqrt{\xi}$ of the theory lies around a few TeV. As we shall see, this situation strikingly improves for freeze-in, where the theory remains valid all the way up to the Planck scale, thanks to  $\xi\lesssim\mathcal{O}(1)$, needed for a successful freeze-in production. This also indicates freeze-in to be a more favourable mechanism of DM production in the present set-up, keeping the high scale validity of the model intact. Before moving on, we would like to mention that there are limits on the DM annihilation cross-section from the non-observation of gamma-ray signals in dwarf satellite galaxies from the MAGIC Cherenkov telescopes and the Fermi Large Area Telescope (LAT)~\cite{MAGIC:2016xys}. However, these bounds typically constraint the low DM mass region $m_X<50$ GeV in our case,
where the direct search bounds are much more severe, hence we do not show them here. It is interesting to note that an effective interaction of the form $\xi\,\widetilde{R}\,\widetilde{X_\mu}\,\widetilde{X^\mu}$ can similarly be written for a fermionic DM $\widetilde{\chi}$: $\frac{\xi}{M_\text{pl}}\,\overline{\widetilde{\chi}}\,\widetilde{\chi}\,\widetilde{R}$ in the Jordan frame. Note that the presence of $1/M_\text{Pl}$ in the non-minimal coupling of the DM $\widetilde{\chi}$ in the Jordan frame brings additional suppression of $1/M_\text{Pl}$ in the coupling strength of the DM in the effective Einstein action. It is thus expected that for fermionic DM, detectable signals will be more suppressed compared to vector DM scenario. For scalar DM $\widetilde{\varphi}$, on the other hand, we can write the well-known interaction $\xi\,\widetilde{\varphi}^2\,\widetilde{R}$ in the Jordan frame. The phenomenology of such non-minimally coupled scalar DM in presence of non-minimally coupled SM Higgs $\xi_h\,\widetilde{H}^\dagger \widetilde{H}\,\widetilde{R}$ has been studied in Ref.~\cite{Ren:2014mta}. In the limit $\xi_h\to 0$ this situation becomes similar to the present framework, where the scalar DM can pair annihilate into the SM final states only via contact interactions $\varphi\varphi\to\text{SM}\,\text{SM}$ in the Einstein frame, opening up the possibility of DM direct detection which is otherwise shown to be absent in Ref.~\cite{Ren:2014mta}.

\section{Freeze-in production of vector dark matter}\label{sec:fiu1x}

Since the effective coupling of the DM with the visible sector in the non-minimally coupled scenario is suppressed by the Planck mass ($\sim\xi/\mpl^2$), hence it is rather natural to assume the DM produced out of equilibrium from the SM bath depending on the choice of the non-minimal coupling $\xi$. In that case, the DM is non-thermally produced in the early Universe via freeze-in~\cite{Hall:2009bx, Bernal:2017kxu}. In this section, we show freeze-in is a viable set-up for the non-minimally coupled vector DM,\footnote{This is in contrast to Ref.~\cite{Kaneta:2021pyx} where the authors have considered a conformally induced Higgs portal.} and the observed relic abundance can be produced for a much lower $\xi$ compared to the freeze-out case, depending on the DM mass and reheating temperature.  

\subsection{Freeze-in via non-minimal coupling}
\label{sec:nonm-fi}

\begin{figure}[htb!]
$$
\includegraphics[scale=0.5]{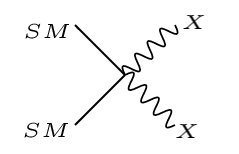}
$$
\caption{Freeze-in production of the abelian vector DM, where SM stands for all the SM particles.}
\label{fig:fimpx}
\end{figure}

In the present scenario, the freeze-in production of the DM occurs through the 2-to-2 scattering of the SM particles in the thermal bath via contact interaction as in Fig.~\ref{fig:fimpx}.
We solve the full BEQ in Eq.~\eqref{eq:BEQ2} numerically with the 2-to-2 annihilation cross-sections collected in Appendix~\ref{sec:ann-fi}. The DM yield, however, can be analytically computed by approximating the annihilation cross-section to be of the form 
\begin{equation}
    \sigma\left(s\right)\sim\frac{\xi^2}{\mpl^4\,m_X^4}\,s^3
\label{eq:sig-msls}
\end{equation}
for a centre-of-mass energy much higher than the DM and SM masses (the general expression is reported in Appendix~\ref{sec:ann-fi}). Note that, the DM is considered to be always massive, and thus the cross-section has a $1/m_X^4$ dependence due to the longitudinal modes of the massive gauge boson $X_\mu$. This gives rise to the DM reaction density
\begin{equation} \label{eq:gamma}
    \gamma \simeq g_a\,g_b
    \frac{\xi^2\,T^{12}}{\mpl^4\,m_X^4}\,,
\end{equation}
where $g_{a,b}$ are the degrees of freedom of the incoming SM particles. The DM yield reads
\begin{equation}
Y_X\left(T\right)\simeq\frac{\xi^2}{\sqrt{g_{\star\rho}}\,g_{\star s}}\,\frac{\left(\Trh^7-T^7\right)}{m_X^4\,\mpl^3}\,,
\label{eq:analy-yld} 
\end{equation}
where assuming an {\it instantaneous} decay for the inflaton, $\Trh$ corresponds to the temperature at which the inflaton decays, and therefore the maximum temperature reached by the thermal bath. We have also neglected the small deviation due to temperature evolution of the numbers of relativistic degrees of freedom. Note that the majority of the DM is produced near the highest temperatures $T \simeq \Trh$ reached by the Universe, which is the characteristic of UV freeze-in~\cite{Hall:2009bx, Elahi:2014fsa, Bernal:2019mhf, Barman:2020plp, Barman:2020ifq}.


Before closing this section, we would like to comment about the instantaneous decay approximation of the inflaton.
While reheating is commonly approximated as an instantaneous event, the decay of the inflaton into SM radiation is a continuous process~\cite{Scherrer:1984fd}.
Away from this approximation for reheating, the bath temperature may rise to a value $\Tmax$ which exceeds $\Trh$~\cite{Giudice:2000ex}.
It is plausible that the DM relic density may be established during this reheating period, in which case its abundance will significantly differ from freeze-in calculations assuming radiation domination.
In particular, it has been observed that if the DM is produced during the transition from matter to radiation domination via an interaction rate that scales like $\gamma(T)\propto T^n$, for $n>12$ the DM abundance is enhanced by a boost factor proportional to $(\Tmax/\Trh)^{n-12}$~\cite{Garcia:2017tuj}, whereas for $n\leq 12$ the difference between the standard UV freeze-in calculation differ 
only by an $\mathcal{O}(1)$ factor from calculations taking into account non-instantaneous reheating.
More recently, it has been highlighted that the critical mass dimension of the operator at which the instantaneous decay approximation breaks down depend on the equation of state $\omega$, or equivalently, to the shape of the inflationary potential at the reheating epoch~\cite{Bernal:2019mhf, Garcia:2020eof, Bernal:2020qyu}.
Therefore, the exponent of the boost factor becomes $(\Tmax/\Trh)^{n-n_c}$ with $n_c \equiv 6+2\, \left ( \frac{3-\omega}{1+\omega} \right ) $, 
showing a strong dependence on the equation of state~\cite{Bernal:2019mhf}.
Subsequent papers have explored the impact of this boost factor in specific models~\cite{Chen:2017kvz, Bernal:2018qlk, Bhattacharyya:2018evo, Chowdhury:2018tzw, Kaneta:2019zgw, Banerjee:2019asa, Chanda:2019xyl, Baules:2019zwk, Dutra:2019xet, Dutra:2019nhh, Mahanta:2019sfo, Cosme:2020mck, Garcia:2020wiy}.
Finally, another way for enhancing the DM abundance occurs in cosmologies where inflation is followed by an epoch dominated by a fluid stiffer than radiation.
In such scenarios, even a small radiation abundance, produced for instance by instantaneous preheating effects, will eventually dominate the total energy density of the Universe without the need for a complete inflaton decay.
In particular, a strong enhancement if DM production happens via interaction rates with temperature dependence higher that $n_c=6$~\cite{Bernal:2020bfj}.

In the present case, as the interaction rate density $\gamma(T) \propto T^{12}$ (cf. Eq.~\eqref{eq:gamma}), a sizeable boost factor is not expected, at least in the standard case where during reheating the inflaton energy density scales like non-relativistic matter. However, as the precise determination of such boost factors depends on the details of the inflationary model (in particular on the energy density carried by the inflaton and its equation-of-state parameter previous to its decay), it is beyond the scope of this study.

\subsection{Gravitational UV freeze-in}\label{sec:min-fi}
In this section we consider the gravitational DM production in the minimal case where $\xi = 0$.
In particular, we have already realised in Sec.~\ref{sec:nonm-fi} that as opposed to the freeze-out scenario, freeze-in supports $\xi\ll 1$ depending on the choice of the DM mass and reheating temperature. However, even if we set $\xi$ to be exactly zero, still gravity can propagate between the DM and the SM once we allow a small fluctuation in the background spacetime. Therefore, in the small $\xi$ limit it is possible that the gravity {\it mediated} interactions may dominate over those due to non-minimal coupling. Specifically, we consider the weak gravity limit of the Eq.~\eqref{eq:abein2_1} and take $\xi=0$. Subsequently, we expand the free part of the Lagrangian around the flat Minkowski background which can be realised via $g_{\mu\nu}=\,\eta_{\mu\nu}+\kappa\,h_{\mu\nu}$, where $h_{\mu\nu}$ is taken to be a small fluctuation over the flat Minkowski spacetime and $\kappa = 1/\mpl$. Due to the smallness of the fluctuation, we allow this perturbative expansion up to the first order in $h_{\mu\nu}$. This also leads us to $g^{\mu\nu} = \eta^{\mu\nu} - \kappa\, h^{\mu\nu}$ and  $\sqrt{-g} \approx 1 + \frac{\kappa\, h}{2}$, where $h = \eta_{\mu\nu}\, h^{\mu\nu}$. Thus we write
\begin{equation}
    \mathcal{L}_\text{gm}(X, {\rm SM})=\,\mathcal{L}_\text{gm}^{(0)}+\,\kappa\, \mathcal{L}_\text{gm}^{(1)}+\mathcal{O}(\kappa^2)+...
\end{equation} 
where $\mathcal{L}_\text{gm}(X, {\rm SM})$ stands for the  Lagrangian of SM matter fields and DM field defined with respect to $g_{\mu\nu}$ while $\xi$ is taken to be zero. Thereafter one can write the perturbed Lagrangian corresponding to all the SM and DM fields up to the leading order of $h_{\mu\nu}$ as\footnote{For a detailed derivation see Refs.~\cite{Choi:1994ax, Holstein:2006bh}} 
\begin{align}
    &\mathcal{L}_{gH}^{(1)}=\,\frac{h}{2}\,(D_\mu H)^\dagger\,(D^\mu H)-\,h^{\mu\nu}\,(D_\mu H)^\dagger\,(D_\nu H)-\,\frac{\kappa\, h}{2}\,V_H\,, \label{lag_per_10}\\
    &\mathcal{L}_{gf}^{(1)}=\,\frac{h}{2}\,(i\,\bar{f}\gamma^\mu\,\partial_\mu f)-\,\frac{i}{2}\,h_{\alpha\beta}\,\bar{f}\,\gamma^\alpha\,\partial^\beta f\,,\\
    &\mathcal{L}_{gX}^{(1)}=\,\frac{1}{2}\,h^{\nu}_{\alpha}\,X_{\mu\nu}\,X^{\mu\alpha}-\,\frac{h}{8}\,X_{\mu\nu}X^{\mu\nu}+\,\bigg(\frac{h\,m_{X}^{2}}{4}\,X_{\alpha}X^{\alpha}-\,\frac{m_{X}^{2}}{2}\,h^{\mu\alpha}\,X_{\mu} X_{\alpha}\bigg). \label{lag_per_1}
\end{align}
Note that all fields and operators in the above set of equations are now contracted with respect to the Minkowski metric $\eta_{\mu\nu}$. 
At this stage we note:
\begin{itemize}
\item The above equations signify that even in the absence of the non-minimal coupling the DM can still be produced from the visible sector via the $s$-channel exchange of massless gravitons as in Fig.~\ref{fig:grav-med}.
\begin{figure}[htb!]
$$
\includegraphics[scale=0.45]{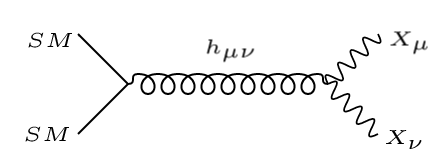}
$$
\caption{Gravitational UV freeze-in production of vector DM via massless graviton mediation.}
\label{fig:grav-med}
\end{figure}
\item  For $\xi \neq 0$, one can similarly employ the metric fluctuation to the terms within the curly bracket of Eq.~\eqref{eq:abein2_1}, which gives rise to coupling strength $\sim\mathcal{O}\left(\xi/\mpl^3\right)$ with gravitational fluctuations ($h_{\mu\nu},\,h$). This coupling strength is more suppressed than the effective non-minimal coupling $\left(\xi/\mpl^2\right)$ that corresponds to the background flat metric ($\eta_{\mu\nu}$). Moreover, this is also suppressed than that due to the metric fluctuations $\left(\sim 1/\mpl\right)$, emerging from the free part of the Lagrangian in Eq.~\eqref{eq:abein2_1}.
This leads us to neglect such coupling consistently in the subsequent analysis. 
\end{itemize}
Therefore, in the presence of both the minimal and non-minimal coupling (for  $\xi\neq 0$), the resulting Lagrangian can be written as
\begin{equation}
    \mathcal{L} \sim \mathcal{L}_{\rm minimal}+\,\mathcal{L}_{\text{non-minimal}}\,,
\end{equation}
with the squared amplitude approximated to be 
\begin{equation}
    \left|\mathcal{M}\right|^2\simeq \left|\mathcal{M}_\text{non-minimal}\right|^2+\left|\mathcal{M}_\text{minimal}\right|^2+\mathcal{O}\left(\xi/\mpl^4\right).    
\end{equation}
We thus separately consider the contributions due to non-minimal coupling and those due to graviton mediation in the present set-up.


\begin{figure}[htb!]
$$
\includegraphics[scale=0.34]{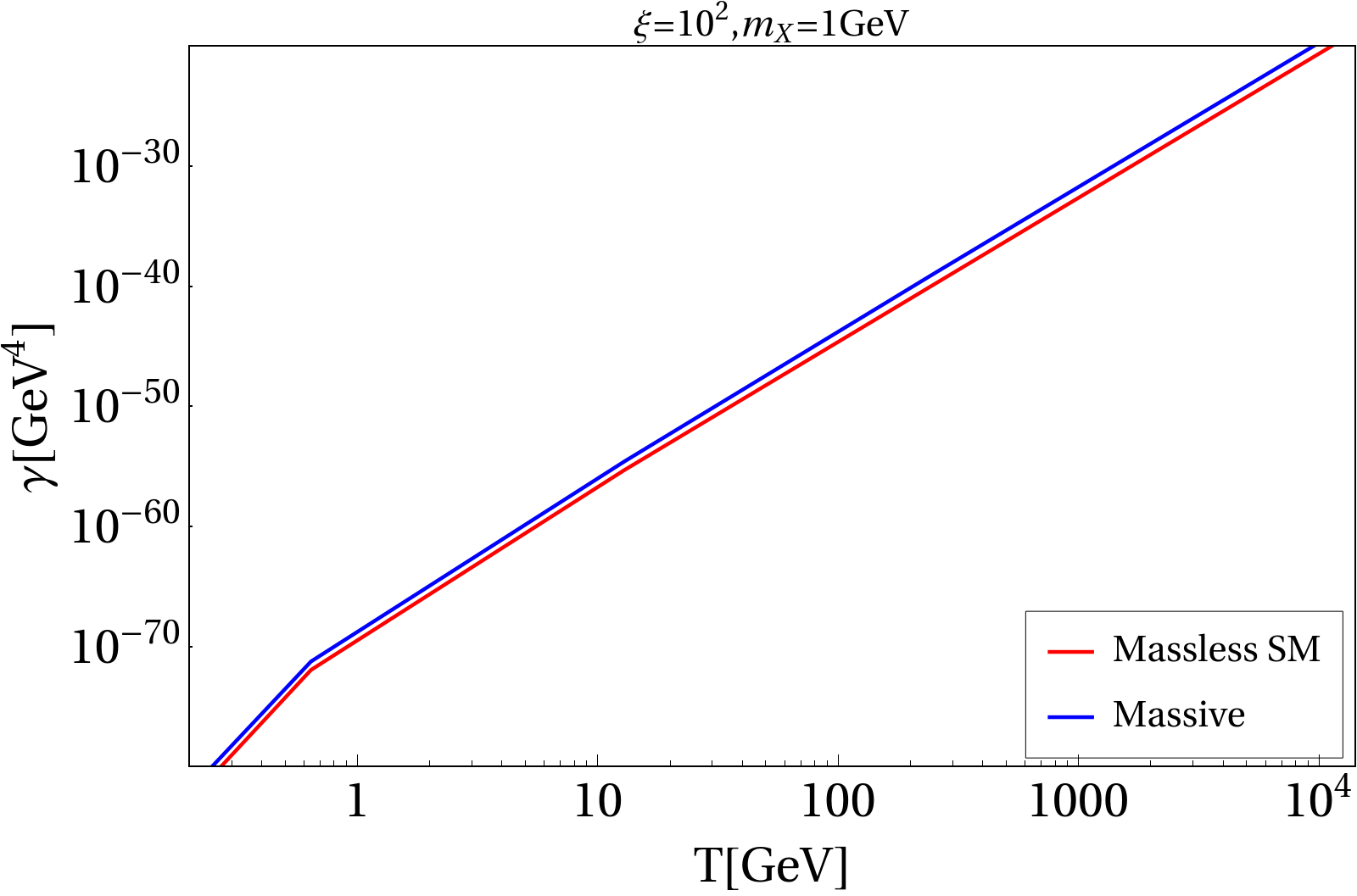}~~~~
\includegraphics[scale=0.34]{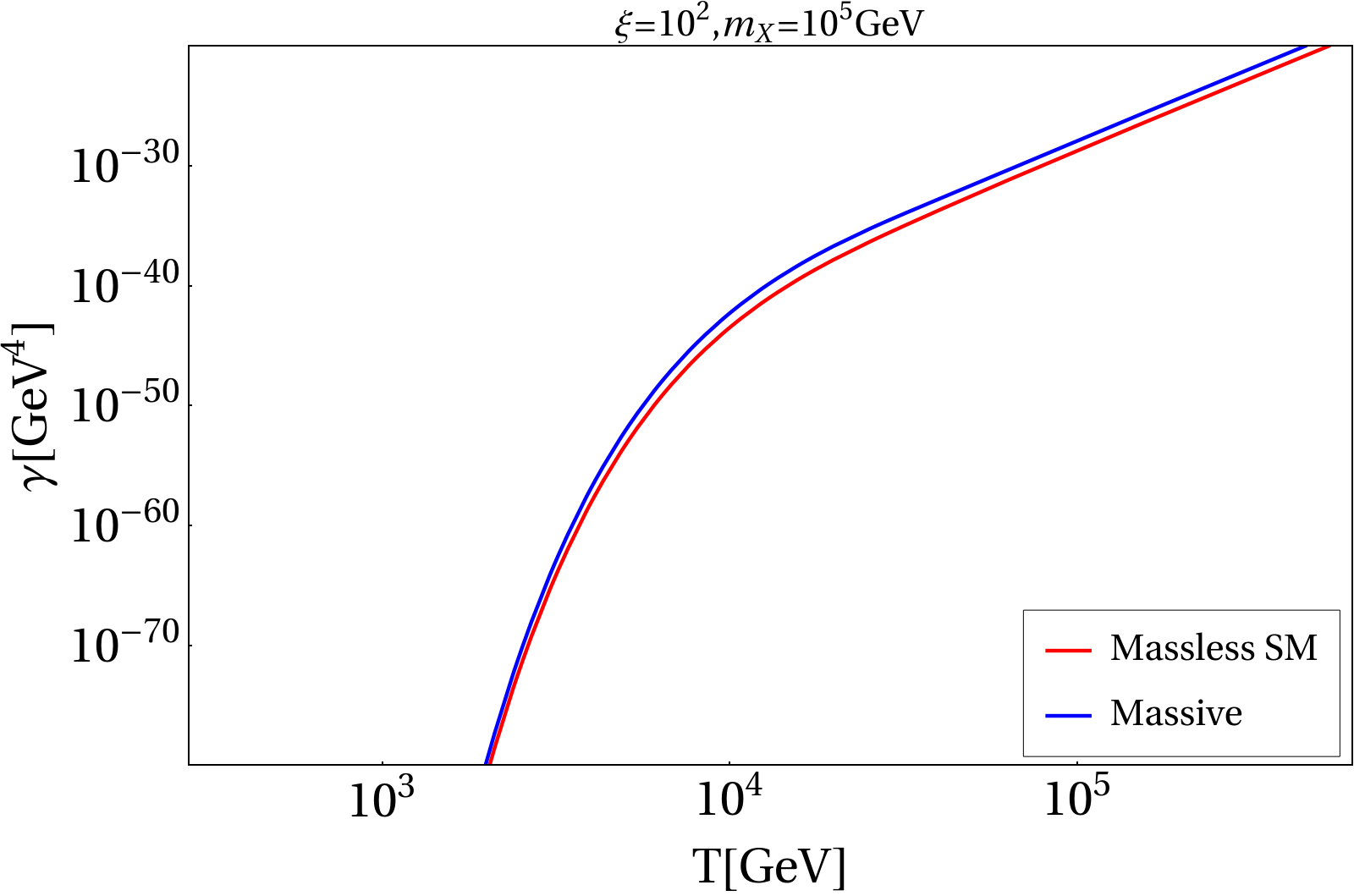}
$$
$$
\includegraphics[scale=0.34]{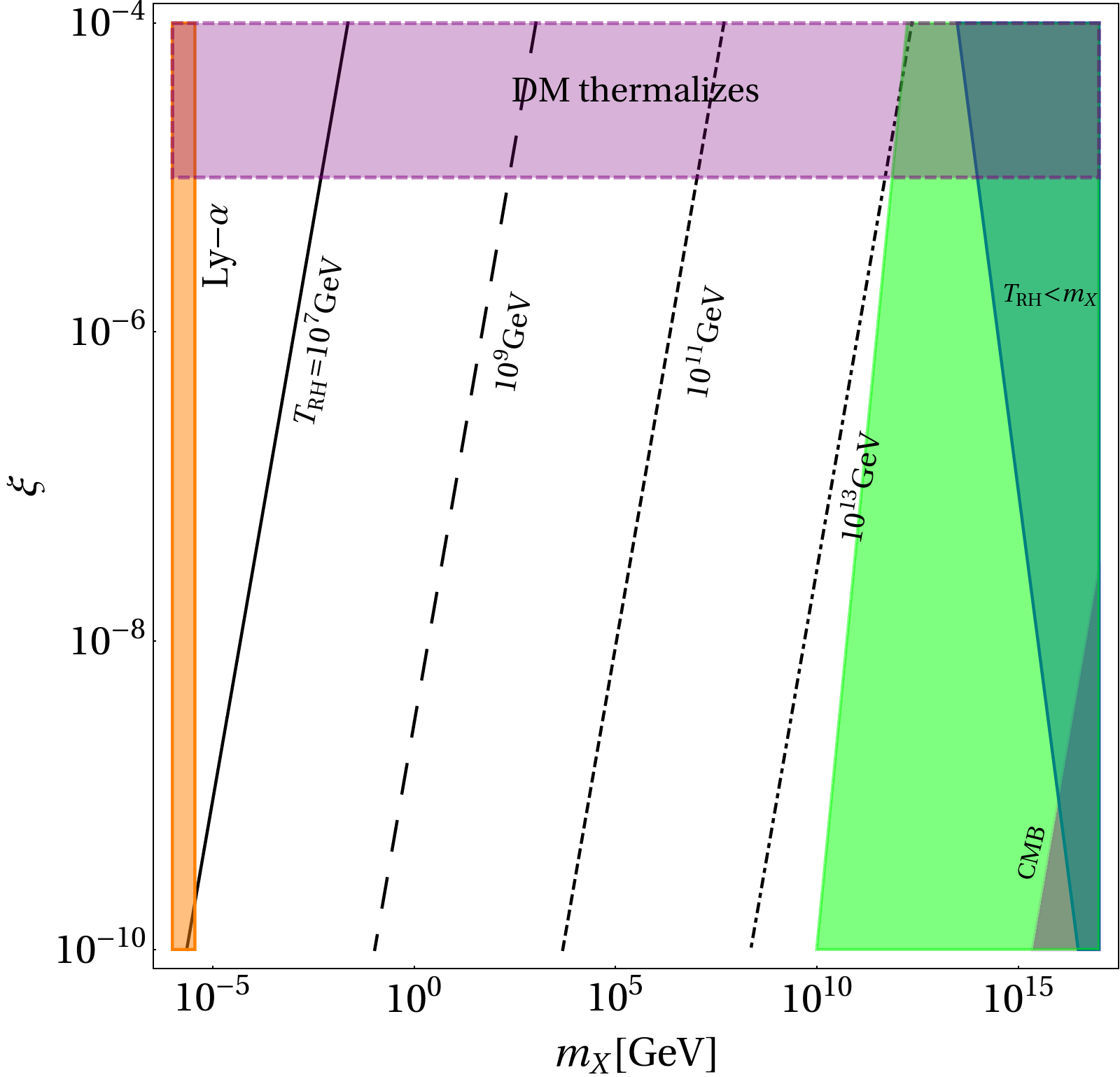}~~~~
\includegraphics[scale=0.34]{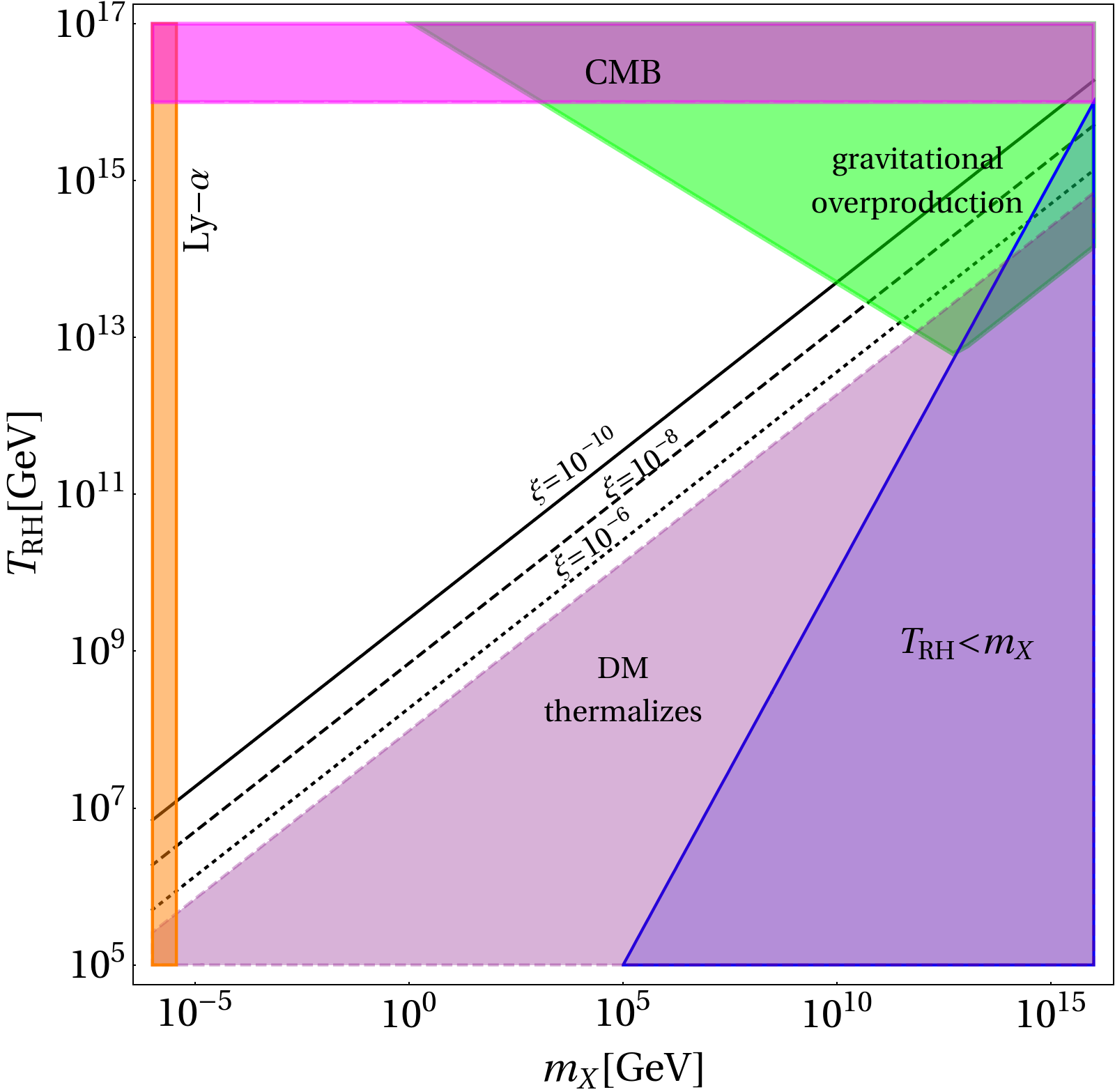}
$$
\caption{Top: Reaction densities as a function of temperature where in red we assume only the SM fields are to be massless, while in blue the SM is assumed to be massive. In either cases the DM is massive with $m_X=10$ GeV (top left) and $m_X=10^5$ GeV (top right). Bottom: The black straight contours correspond to the central value of Planck observed DM relic abundance for different choices of $\Trh$ (shown by different patterns) in the left panel and for different choices of $\xi$ in the right panel. The shaded regions are disallowed from DM thermalization: $\xi \gtrsim 10^5$ (purple), CMB: $\Trh>10^{16}$~GeV (magenta), DM production in instantaneous approximation  $\Trh < m_X$ (blue), Lyman-$\alpha$ limit: $m_X \lesssim 3.5$~keV (orange) and DM overproduction due to $s$-channel graviton mediated process (green).
}
\label{fig:rel-fi}
\end{figure}

The interaction rate density for such 2-to-2 graviton mediated process reads~\cite{Garny:2015sjg, Tang:2017hvq, Garny:2017kha, Bernal:2018qlk, Barman:2021ugy, Bernal:2021qrl}
\begin{equation}
    \gamma(T) = \alpha\, \frac{T^8}{\mpl^4}\,,
\end{equation}
with $\alpha \simeq 2.3\times 10^{-3}$. For $T \ll \Trh$, one can analytically obtain the DM yield at the end of reheating as
\begin{equation} \label{eq:FIlight}
    Y_X(T_0) \simeq \frac{45\, \alpha}{2\pi^3\, \gss} \sqrt{\frac{10}{\gs}} \left(\frac{\Trh}{\mpl}\right)^3,
\end{equation}
in the case $m_X \ll \Trh$.%
\footnote{Two comments are in order.
Firstly, we note that the DM abundance could also be set entirely in the hidden sector by the dark freeze-out of 4-to-2 interactions, where four DM particles annihilate into two of them~\cite{Carlson:1992fn, Bernal:2015xba, Bernal:2017mqb, Bernal:2020gzm, Barman:2021ugy}.
However, such a possibility is sub-dominant due to a strong suppression by higher orders of $\mpl$.
Secondly, the gravitational production can be enhanced in scenarios with extra dimensions, see e.g., Refs.~\cite{Lee:2013bua, Lee:2014caa, Han:2015cty, Rueter:2017nbk, Kim:2017mtc, Rizzo:2018ntg, Carrillo-Monteverde:2018phy, Kim:2018xsp, Rizzo:2018joy, Goudelis:2018xqi, Brax:2019koq, Folgado:2019sgz, Goyal:2019vsw, Folgado:2019gie, Kang:2020huh, Chivukula:2020hvi, Kang:2020yul, Kang:2020afi, Bernal:2020fvw, Bernal:2020yqg, deGiorgi:2021xvm}.}

In the top panel of Fig.~\ref{fig:rel-fi} we illustrate a comparison of the reaction densities $(\gamma)$ considering only the SM particles to be massless (in red) with the one where all states (SM and DM) are massive (in blue), for two different DM masses: 1 GeV (left) and $10^5$ GeV (right). We see the two scenarios behave identically with temperature, irrespective of the DM mass, with a difference in magnitude only in the percentage level. Thus, the massless SM approximation is a valid one in the present scenario. The DM, however, is considered always to be massive as mentioned before.

In the bottom panel of Fig.~\ref{fig:rel-fi} we illustrate the parameter space matching the whole observed DM relic density for the freeze-in scenario. In both the panels, the black straight lines indicate contours satisfying the observed relic density. From the left panel we see, for a smaller choice of $\Trh$, the observed DM abundance can be obtained for lighter DM for a fixed $\xi$.  This is understandable, since the DM abundance varies as $\Omega_X h^2 \propto \xi^2\, \Trh^7/m_X^3$, hence a larger $\Trh$ calls for a heavier DM mass to produce the right abundance for a fixed $\xi$. Thus, the contours are in increasing order of $\Trh$ from left to right. This can also be verified from the right panel plot where the black straight line contours corresponding to right DM abundance are in increasing order of $\xi$ from left to right. Notice, in the freeze-in framework it is possible to have DM mass from keV up to $\sim \mathcal{O}(10^{10})$~GeV satisfying the observed abundance, unlike the freeze-out case where the DM mass can be maximum $\sim 55$~TeV. The contribution from the gravitational UV freeze-in is shown in green, where the coloured region depicts DM overproduction when the DM is minimally coupled to the SM. In the left panel we project this bound in the $\xi$-$T_\text{RH}$ plane, where we see DM can be gravitationally over produced for $m_X\gtrsim 10^{10}$ GeV and the production becomes comparable to that via non-minimal coupling for $T_\text{RH}\gtrsim 10^{13}$ GeV. This is also reflected in the right panel, where we find DM production from gravitational UV freeze-in overwhelms the production due to non-minimal coupling for heavier DM and large $T_\text{RH}$ (cf. Eq.~\eqref{eq:FIlight}).    
The relic density allowed parameter space for the DM can further be constrained by several bounds as shown by the coloured regions in the bottom right panel. Here we summarise them. First of all, it is important to note that one can not take $\xi$ arbitrarily large since in that case the reaction rate of DM production may exceed the Hubble rate at large temperatures, making the DM thermal. We find the condition $\Gamma_\text{int}=\langle\sigma v\rangle_i n^i_\text{eq}<H$ (where $i\in\text{SM}$) can be satisfied with $\xi\lesssim 10^{-5}$. For heavier DM $m_X\gtrsim 10^5$ GeV, however, this condition is somewhat relaxed as the thermally averaged interaction cross-section has  $\langle\sigma v\rangle\propto 1/m_X^4$ dependence. We thus project a rather conservative bound on $\xi$ from the non-thermal condition. A major part of the viable parameter space is disallowed from the instantaneous inflaton decay approximation, which does not allow to have a DM mass larger than the reheating temperature $\Trh$. This is shown by the blue region. The upper limit on the inflationary scale is constrained from CMB measurements~\cite{Planck:2018jri}: $H_I^\text{CMB}\leq 2.5\times 10^{-5}\mpl$, which in turn allows $\Trh\leq 10^{16}$ GeV. DM mass below 3.5~keV is forbidden from the measurements of the free-streaming of warm DM from Lyman-$\alpha$ flux-power spectra~\cite{Irsic:2017ixq, Ballesteros:2020adh, DEramo:2020gpr}. This is denoted by the thin orange region. Finally, $\Trh\lesssim 4$ MeV can potentially disturb the measurement of light element yield from big bang nucleosynthesis (BBN) and hence such a small reheating temperature can not be allowed.
\section{Conclusions}\label{sec:concl}
The gravitational interaction of DM with the SM fields is inevitable, and supported by all astrophysical and cosmological evidences for the existence of DM.
This leads us to consider a scenario where a vector boson DM couples to the SM fields through an unique dimension-4 operator, connecting a pair of DM fields (i.e., $\widetilde{X}_\mu \widetilde{X}^\mu$) with the scalar curvature of the background spacetime ($\widetilde{R}$) via a non-minimal coupling.
Unlike most of the cases studied in the literature, here we consider {\it only} the DM fields are non-minimally coupled to gravity. 
This gives rise to the non-minimal coupling of DM to the visible sector in the Einstein frame leading to DM pair annihilation (production) to (from) the SM particles.
In all such cases, the effective DM-SM interaction strength is parametrized by the coupling $\sim\xi/\mpl^2$, that can further be constrained by different theoretical and experimental bounds as discussed. We take up  the simplest form of non-minimally coupled abelian vector DM, where the DM is considered to be $\mathbb{Z}_2$-odd (while all the SM fields are $\mathbb{Z}_2$-even) and owns a Stuecklberg mass term. This helps in reducing the number of free parameters for the theory to only two: the non-minimal coupling $\xi$ and the DM mass $m_X$, which in turn portrays the minimalistic feature of this model. 
By considering the DM to be a weakly interacting massive particle (WIMP), we find, to produce the observed relic abundance via freeze-out, $\xi$ turns out to be $\sim\mathcal{O}\left(10^{30}\right)$ for DM mass $m_X \lesssim 55$~TeV such that perturbative unitarity is not violated. We also show, for such choice of the non-minimal coupling, it is possible to evade strong spin-independent direct detection bounds arising from PandaX-4T experiment (or even future projection of XENONnT), however a large part of the viable parameter space also gets submerged into the so-called neutrino floor. Thus, the model provides testability for such a gravitationally coupled simple WIMP scenario in DM scattering experiments. 

In the present set-up, it is also interesting to address the freeze-in production of the vector DM, where only 2-to-2 annihilation of the bath particles can give rise to the required DM abundance. The freeze-in mechanism turns out to be more preferable in this scenario since in that case the non-minimal coupling turns out to be $\mathcal{O}(1)$, that in turn keeps the theory valid till the Planck scale. The freeze-in production can occur both via the non-minimal coupling to gravity, as well as via $s$-channel graviton exchange, where the latter is present even in the limit $\xi\to 0$. We notice that the freeze-in yield replicates the typical UV nature, where the DM number density reaches maximum at the highest temperature of the thermal bath (which is the reheating temperature, assuming instantaneous inflaton decay), and immediately saturates. Contrary to the freeze-out scenario, a much smaller $\xi$ is require to obtain the Planck observed relic density, depending on the choice of the reheating temperature and satisfying several bounds arising from perturbative unitarity, warm DM limit, scale of inflation and BBN.

\section*{Acknowledgement}
The authors would like to thank Óscar Catà, Kunio Kaneta, Tanmoy Paul, Sabir Ramazanov, Jing Ren, Javier Rubio and Soumitra Sengupta for many useful correspondences and fruitful comments. BB would like to thank Alexander Pukhov for helping with the freeze-in calculations in {\tt MicrOmegas}. We would like to acknowledge the anonymous referee for raising many relevant points that helped in improving the manuscript. BB and NB received funding from the Patrimonio Autónomo - Fondo Nacional de Financiamiento para la Ciencia, la Tecnología y la Innovación Francisco José de Caldas (MinCiencias - Colombia) grant 80740-465-2020.
NB also received funding from the Spanish FEDER/MCIU-AEI under grant FPA2017-84543-P.
This project has received funding /support from the European Union's Horizon 2020 research and innovation programme under the Marie Skłodowska-Curie grant agreement No 860881-HIDDeN. 
\appendix
\section{Transformation from Jordan frame to Einstein frame}\label{sec:jor-ein}
The action in Jordan frame can be written as follows:
\begin{equation}
S=\,\int d^4x\,\sqrt{-\widetilde{g}}\Bigg[\frac{1}{2}\bigg(\mpl^2-\xi\,\widetilde{X}_{\mu}\widetilde{X}^{\mu}\bigg)\,\widetilde{R}+\,\widetilde{\mathcal{L}}_\text{DM}+\,\widetilde{\mathcal{L}}_\text{SM}\Bigg]\label{eq:S1} 
\end{equation}
where $\widetilde{g}_{\mu\nu}$ stands for the metric in the Jordan frame while  $\widetilde{\mathcal{L}}_\text{SM}$ and $\widetilde{\mathcal{L}}_\text{DM}$ represent the Lagrangian for the SM and the dark sector respectively, and can be written in explicit form as follows,
\begin{equation}\begin{aligned}
& \mathcal{\widetilde{L}}_\text{DM} = -\frac{1}{4}\widetilde{X}_{\mu\nu} \widetilde{X}^{\mu\nu} + \frac{1}{2}m_X^2 \widetilde{X}_\mu \widetilde{X}^\mu.
    \end{aligned}
\end{equation}
\begin{equation}\begin{aligned}
& \mathcal{\widetilde{L}}_\text{SM} = \widetilde{g}\,^{\mu\nu}\,(\widetilde{D}_{\mu} \widetilde{H})^{\dagger}\,(\widetilde{D}_{\nu} \widetilde{H})-\frac{1}{4}\widetilde{F}_{\mu\nu} \widetilde{F}^{\mu\nu} + \frac{i}{2}\,\,\widetilde{\overline{f}}\,\overleftrightarrow{\widetilde{\slashed{\nabla}}}\,\widetilde{f}+\,\widetilde{\mathcal{L}}_{Y}.
    \end{aligned}
\end{equation}
As stated earlier, we follow the metric convention $\eta_{\mu\nu}=\,\{+,-,-,-\}$ and consider the conformal transformation as
\begin{equation}
g_{\mu\nu}=\,\omega^2\,\widetilde{g}_{\mu\nu}
\end{equation}
where $g_{\mu\nu}$ stand for the spacetime metric of Einstein frame and $\omega$ is known to be the conformal factor. We identify
\begin{equation}
    1-\frac{\xi\,\widetilde{X}_{\mu}\widetilde{X}^{\mu}}{\mpl^2} \equiv \omega^2.
\end{equation}
Note that all the un-tilde quantities belong to the Einstein frame. 
We mention that the spacetime coordinates are not altered due to the conformal transformation. Therefore the ordinary derivative $\widetilde{\partial}=\partial$, whereas the covariant derivative $\widetilde{\nabla}_{\mu}$ is defined with respect to the $\widetilde{g}_{\mu\nu}$.    It can also be perceived that when covariant derivative operates on the scalar it reduces to ordinary derivative and thus we write: $\widetilde{\nabla}_{\mu}\phi=\,\widetilde{\partial}_\mu\phi=\partial_\mu \phi$. Furthermore note that  $\widetilde{H},\,\widetilde{f},\,\widetilde{X}_\mu$ and $\widetilde{F}_{\mu\nu}$ all remain unaffected by the conformal transformation {\it i.e.}, $\widetilde{H}=H,\,\widetilde{f}=f,\,\widetilde{X}_\mu=\,X_\mu$ and $\widetilde{F}_{\mu\nu}=\,F_{\mu\nu}$.  On the other hand,  $\widetilde{F}^{\mu\nu}=\,\widetilde{g}^{\alpha\mu}\,\widetilde{g}^{\beta\nu}\,\widetilde{F}_{\alpha\beta}=\omega^4\,F^{\mu\nu}$.
In the fermionic sector $\widetilde{\slashed{\nabla}}=\,\gamma^\mu\,\widetilde{\nabla}_\mu$, where $\gamma^{\mu}$ are the gamma matrices in the Jordan frame and can be connected to the Einstein frame by using the vierbein.
Now we analyse term by term of the action in Eq.~(\ref{eq:S1}). $\widetilde{R}$ is the Ricci scalar in the Jordan frame and can be related to the Ricci scalar in the Einstein frame as below~\cite{Carroll:2004st}
\begin{equation}
\widetilde{R} = \omega^2 \left[R - 6\, g^{\alpha\beta} \nabla_{\alpha} \nabla_{\beta} ({\rm ln}\,\omega) + 6\, g^{\alpha\beta} \nabla_{\alpha} ({\rm ln}\, \omega)\, \nabla_{\beta} ({\rm ln}\,\omega)\right].
\label{ricci_transformation}
\end{equation}
Using the above relation we replace $\widetilde{R}$ in the Eq.~(\ref{eq:S1}). We also transform all the parameters associated with the term $\widetilde{R}$ to Einstein frame by using the appropriate transformation relations. Thus the first two terms in Eq.~\eqref{eq:S1} turn out to be
\begin{equation}
    \int d^4x\, \sqrt{-g} \left[\frac{\mpl^2\,R}{2} + \frac{3 \omega^4}{4\, \mpl^{2}}\, \nabla_{\alpha} (\xi\, X_{\mu}X^{\mu})\, \nabla^{\alpha} (\xi\, X_{\mu}X^{\mu})\right].
\label{ricci_ein_1}
\end{equation}

Following the discussion above Eq.~(\ref{ricci_transformation}), one obtains
\begin{equation}
    \widetilde{X}_{\mu\nu}\widetilde{X}^{\mu\nu}=\,\tilde{g}^{\alpha\mu}\tilde{g}^{\beta\nu}\,X_{\mu\nu}\,X_{\alpha\beta}=\,\omega^4\,g^{\alpha\mu}g^{\beta\nu}\,X_{\mu\nu}X_{\alpha\beta}=\,\omega^4\,X_{\mu\nu}X^{\mu\nu}
\label{gauge_dm_transformation}
\end{equation}
\begin{equation}
    \widetilde{X}_{\mu} \widetilde{X}^{\mu}=\,\widetilde{g}^{\mu\nu}\,X_{\mu}X_{\nu}=\,\omega^2\,g^{\mu\nu}\,X_{\mu}X_{\nu}\label{dm_transformation}
\end{equation}
and
\begin{equation}
    \widetilde{g}\,^{\mu\nu}\,(\widetilde{D}_{\mu} \widetilde{H})^{\dagger}\,(\widetilde{D}_{\nu} \widetilde{H})\,=\, \omega^2\,g^{\mu\nu}\,(D_{\mu} H)^{\dagger}\,(D_{\nu} H).
\label{higgs_transformation}
\end{equation}

For fermions some comments are in order: 
\begin{itemize}
\item
 In case of fermions, the metrics of the two frames are connected by the vierbein.
 \begin{equation}
 \widetilde{g}_{\mu\nu}=\,e_{\mu}^{a}\,e_{\nu}^{b}\,\eta_{ab}\,\,\,\,\,\,\,\,\,\,\,\,\,\,\,\,\,\,\,\,\,{\rm det}(e_{\nu}^{q})=\,\sqrt{-\widetilde{g}}\,\,\,\,\,\,\,\,\,\,\,\,\,\,\,\,\,\,\,\,\,e_{\mu}^{a}\,e_{b}^{\mu}=\delta_{b}^{a}.
 \end{equation}
Furthermore, the vierbein satisfy: $e_{\mu}^{a}=\,\omega^{-1}\,\delta_{\mu}^{a}$ \,\,\,and\,\,\, $e^{\mu}_{a}=\,\omega\,\delta^{\mu}_{a}$. 
\item
In the case of the Einstein frame ($g_{\mu\nu}$) we take the background to be flat and consider $g_{\mu\nu} \to \,\eta_{\mu\nu}$. Consequently the conformal transformation becomes: $\tilde{g}_{\mu\nu}=\, \omega^{-2}\,\,\eta_{\mu\nu}$~\cite{Ren:2014mta}. 
\item
Vierbein depicts the connection between two frames and its two indices such as $(\mu$, $\nu$, $\alpha...)$ represent the indices for Jordan frame and $(a,\, b,\, c,...)$ stand for Einstein (flat) frame. For example, one can see below (Eq.~(\ref{partial_fermion_tr})) that how $\gamma^\mu$ and $\gamma^a$ are connected by the vierbein. 
\item 
All indices of Jordan and flat frame are contracted with the corresponding metric such as $\tilde{g}_{\mu\nu}$ and $\eta_{\mu\nu}$, respectively. 
\end{itemize}
Using this relation and following the discussion above Eq.~(\ref{eq:S1}), we get partially transformed fermionic action as follows:
 \begin{eqnarray}
 \widetilde{S}_{f}&=&\,\int d^4x\,\, {\rm det}\,(e_{\nu}^{q})\,\bigg[\frac{i}{2}\,\bar{f}\gamma^\mu\,\overleftrightarrow{\widetilde{\nabla}}_{\mu}\,f\bigg]\nonumber\\
 &=&\,\int d^4x \,\,\omega^{-4}\sqrt{-g}\,\bigg[\frac{i}{2}\,\bar{f}\gamma^a\,e_{a}^{\mu}\,\overleftrightarrow{\widetilde{\nabla}}_{\mu}\,f\bigg]
 \label{partial_fermion_tr}\\
 &=&\,\int d^4x \,\,\omega^{-4}\sqrt{-g}\,\bigg[\frac{i}{2}\,\bar{f}\gamma^a\,\omega\,\delta_{a}^{\mu}\,\overleftrightarrow{\nabla_{\mu}}\,f\bigg].
 \label{partial_fermion_1}
 \end{eqnarray}
In case of fermion the covariant derivative is defined as: $\nabla_\mu\,f=\,\partial_\mu\,f+\frac{i}{2}\,\omega_{\mu}\,^{mn}\,\sigma_{mn}$, where $\omega_{\mu}\,^{mn}$ are the antisymmetric coefficients of the spinor connection and $\sigma_{mn}=\,\frac{i}{2}\,(\gamma_{m}\gamma_{n}-\gamma_{n}\gamma_{m})$. Thus putting this in the Eq.~(\ref{partial_fermion_1}) we get, 
\begin{equation}
S_{f} = \,\int d^4x \,\,\frac{i\,\sqrt{-g}}{2\,\omega^3}\,\bar{f}\,\gamma^a\,\overleftrightarrow{\partial_a}\,f\,-\,\frac{1}{4}\,\int\,d^4x \frac{\sqrt{-g}}{\omega^3}\,\bar{f}\,\gamma^a\,\delta^{\mu}_{a}\,\omega_{\mu}\,^{mn}\,\sigma_{mn}\,f\,.
\label{partial_fermion_2}
\end{equation}
At this stage, let us elaborately analyse the term $\omega_{\mu}\,^{mn}$ as below:
\begin{eqnarray}
\omega_{\mu}\,^{mn}&=&\,e_{\nu}^{m}\,\Gamma^{\nu}_{\sigma\mu}\,e^{\sigma  n}-\,(\partial_{\mu}\,e^{m}_{\nu})\,e^{\nu n}\nonumber\\
&=&\,\frac{\omega^2}{2}\,\delta_{\nu}^{m}\,\eta^{ab}\,\delta_{a}^{\nu}\,\delta_{b}^{\beta}\,\bigg[\partial_\mu(\omega^{-2}\eta_{\sigma \beta})+\,\partial_\sigma(\omega^{-2}\eta_{\mu \beta})-\,\partial_\beta(\omega^{-2}\eta_{\sigma \mu})\bigg]\,\eta^{\sigma \alpha}\,\delta_{\alpha}^{n}\nonumber\\
&+&\,\frac{\partial_{\mu}\,\omega}{\omega}\,\delta_{\nu}^{m}\,\eta^{\nu \alpha}\,\delta_{\alpha}^{n}.
\label{spin_connection_1}
\end{eqnarray}
Further simplifying the above, we obtain
\begin{equation}
    \omega_{\mu}\,^{mn}=\,-\,\frac{1}{\omega}\,(\delta_{\mu}^{m}\,\partial^n \omega\,-\,\delta_{\mu}^{n}\,\partial^m\,\omega).
\label{spin_connection_2}
\end{equation}
Let us put Eq.~(\ref{spin_connection_2}) in the second term of Eq.~(\ref{partial_fermion_2}) and use the following properties of gamma matrices:
\begin{equation}
    \left\{\gamma^m,\,\gamma^n\right\}=\,2\,\eta^{mn}\,\,\,\,\,\,\,\,\,\,\,\,\,\,\,\,\,\,\,\gamma_m\,\gamma^m=\,4\,I
\end{equation}
Therefore, the second term of Eq.~(\ref{partial_fermion_2}) becomes:
\begin{eqnarray}
    &&-\frac{1}{4}\,\int\,d^4x \,\frac{\sqrt{-g}}{\omega^3}\,\bar{f}\,\gamma^a\,\delta^{\mu}_{a}\,\omega_{\mu}\,^{mn}\,\sigma_{mn}\,f \nonumber \\
    &&\qquad =\frac{i}{8}\,\int\,d^4x\, \sqrt{-g}\,\bar{f}\,\gamma^p\,\delta_{p}^{\mu}\,(\delta_{\mu}^{m}\,\partial^n\,\omega-\,\delta_{\mu}^{n}\,\partial^m\,\omega)\,(\gamma_{m}\gamma_n-\gamma_n\gamma_m)\,f\nonumber\\
    &&\qquad =\,\,\frac{3\,i}{2}\,\int\,d^4x\,\frac{\sqrt{-g}}{\omega^4}\,\bar{f}\,(\slashed{\partial}\,\omega)\,f=\,\frac{3\,i}{2}\,\int\,d^4x\,\frac{\sqrt{-g}}{\omega^4}\,\bar{f}\,(\partial^m \,\omega)\,\gamma_m\,f\,.
    \label{fermion_ein_1}
\end{eqnarray}
Therefore Eq.~(\ref{partial_fermion_2}) becomes: 
\begin{equation}
S_{f}=\,\int\,d^4x\, \sqrt{-g}\,\bigg[\frac{i}{\omega^3}\,\bar{f}\,\gamma^a\,\partial_a\,f+\,\frac{3\,i}{\omega^4}\,\bar{f}\,(\slashed{\partial}\,\omega)\,f\bigg].
\label{final_ein_fermion}
\end{equation}
Note: $\omega_{\mu}\,^{mn}$ contains $\partial$ operators which also possess over left right arrow as similar to the $\partial$ operator in the first term of Eq.~(\ref{partial_fermion_2}). 
Now Eq.~(\ref{final_ein_fermion}), has been written by removing the over left right arrow from the $\partial$ operators and consequently a $1/2$ factor is removed from both the terms.
Additionally, the SM gauge field sector will exactly behave as the dark sector under conformal transformation such as shown in Eq.~(\ref{gauge_dm_transformation}).
Therefore combining Eqs.~(\ref{ricci_ein_1}), (\ref{gauge_dm_transformation}), (\ref{gauge_dm_transformation}) (\ref{dm_transformation}), (\ref{higgs_transformation}) and (\ref{final_ein_fermion}) we get the final form of action in the Einstein frame for the whole setup as follows:
\begin{align}
\mathcal{S}= \int d^4x \sqrt{-g}&\Biggl[\frac{\mpl^2\,R}{2}+\,\frac{3\,\omega^4}{4\,\mpl^{2}}\,\nabla_{\alpha}(\xi\,X_{\mu}X^{\mu})\,\nabla^{\alpha}(\xi\,X_{\mu}X^{\mu})-\frac{1}{4}\,X_{\mu\nu}\,X^{\mu\nu}+\frac{1}{2\omega^2}m_X^2\,X_\mu\,X^\mu\nonumber\\
&+\,\frac{1}{\omega^4}\,(\mathcal{L}_{Y}-V(H))+\,\frac{1}{\omega^2}(D_{\mu} H)^{\dagger}(D^{\mu} H)
+\,\frac{i}{\omega^3}\,\bar{f}\,\gamma^\mu\,\partial_\mu\,f\nonumber\\
&-\,\frac{1}{4}\,g^{\mu\nu}\,g^{\lambda\rho}F_{\mu\nu}^{(a)}\,F_{\nu\rho}^{(a)}+\,\frac{3\,i}{\omega^4}\,\bar{f}\,(\slashed{\partial}\,\omega)\,f\Biggr],
\end{align}
where $V\left(H\right)$ is the renormalizable Higgs potential for the SM. Here $\mathcal{L}_Y$ contains the Yukawa interaction terms. 

\section{Interactions with metric fluctuation }\label{metric:fluctuation}
Consider the following Lagrangian:
\begin{equation}
\mathcal{L}_\text{gm}(X, {\rm SM})=\,\mathcal{L}_\text{gm}^{(0)}+\,\kappa\, \mathcal{L}_\text{gm}^{(1)}+\mathcal{O}(\kappa^2)+...
\end{equation}
where $\mathcal{L}_\text{gm}(X, {\rm SM})$ stands for the Lagrangian of all the matter fields, including the SM and the DM, defined with respect to $g_{\mu\nu}$. This can be written as
\begin{align}
&\mathcal{L}_{gH}=\,\sqrt{-g}\,\big[\,g^{\mu\nu}(D_\mu H)^\dagger\,(D_\nu H)-\,V_H\big]\\
&\mathcal{L}_{gA}=\,\sqrt{-g}\,\bigg[-\frac{1}{4}\,F_{\mu\nu}\,F^{\mu\nu}\bigg]\\
&\mathcal{L}_{gf}=\,\sqrt{-g}\,\big[\,i\,\bar{f}\gamma^{\alpha}\partial_\alpha\,f\big]\\
&\mathcal{L}_{gX}=\,\sqrt{-g}\,\bigg[-\frac{1}{4}\,X_{\mu\nu}X^{\mu\nu}+\,\frac{1}{2}\,m_{X}^{2}X_{\mu}\,X^\mu\bigg]. 
\end{align}
$\mathcal{L}_\text{gm}^{(0)}$ depicts the  Lagrangian of the SM and DM fields in the Minkowski ($\eta_{\mu\nu}$) spacetime.  We refer the readers to~\cite{Choi:1994ax} for detail derivation of the interaction terms. For illustration purposes, here we analyse $\mathcal{L}_{gH}$ under the metric fluctuation: $g_{\mu\nu}=\,\eta_{\mu\nu}+\kappa\,h_{\mu\nu}$, where due to the smallness of the fluctuation $(h_{\mu\nu})$, we allow the perturbative expansion of the Lagrangian up to the first order in $h_{\mu\nu}$.
This also leads us to $g^{\mu\nu}=\,\eta^{\mu\nu}-\,\kappa\,h^{\mu\nu}$ and  $\sqrt{-g}\approx\,1+\frac{\kappa\,h}{2}$, where $h=\,\eta_{\mu\nu}\,h^{\mu\nu}$. 
Using the metric fluctuation and allowing up to the leading order in $h_{\mu\nu}$,  $\mathcal{L}_{gH}$ can be written as
\begin{eqnarray}
\mathcal{L}_{gH} &&= \bigg(1+\frac{\kappa\,h}{2}\bigg)\,\bigg[(\eta^{\mu\nu}-\,\kappa\,h^{\mu\nu} )\,(D_\mu H)^\dagger\,(D_\nu H)-\,V_H\bigg]\nonumber\\
&&=\big[\eta^{\mu\nu}\,(D_\mu H)^\dagger\,(D_\nu H)-V_H\big]+\frac{\kappa\,h}{2}(D_\mu H)^\dagger\,(D^\mu H)\nonumber\\
&&\quad -\kappa\,h^{\mu\nu}\,(D_\mu H)^\dagger\,(D_\nu H)
-\frac{\kappa\,h}{2}\,V_H\,
\label{higgs_fluc_1}
\end{eqnarray}
from which one can find the relevant interaction terms.

\section{Annihilation cross-section for freeze-in}\label{sec:ann-fi}
We have used {\tt CalcHEP}~\cite{Belyaev:2012qa} to calculate the cross-sections for 2-to-2 processes with a pair of DM in the final state as a function of the CM energy where $f,V$ and $H$ stand respectively for the SM fermions, SM gauge bosons (massive) and the SM Higgs.
\begin{align}
\sigma\left(s\right)_{ff\to XX} =& \frac{N_c\,\xi^2\,(4\,m_X^2-s)}{1440\,\mpl^4\,m_X^4\,\pi\,s^6}\Biggl[1152m_f^8 \left(s-4 m_X^2\right)^4\nonumber\\
&- 64m_f^6 s \left(s-4 m_X^2\right)^2 \left(288 m_X^4-94 m_X^2 s+13 s^2\right)\nonumber\\
&+32m_f^4 s^2 \left(3456 m_X^8-2416 m_X^6 s+771 m_X^4 s^2-86 m_X^2 s^3+6 s^4\right)\nonumber\\
&-2m_f^2 s^3 \left(9216 m_X^8-5696 m_X^6 s+1596 m_X^4 s^2-236 m_X^2 s^3+s^4\right)\nonumber\\
&-3 s^4 \left(-384 m_X^8+224 m_X^6 s+96 m_X^4 s^2-6 m_X^2 s^3+s^4\right)\Biggr]\\
\sigma\left(s\right)_{VV\to XX } =& \frac{\xi^2\,(s-4\,m_X^2)}{4320\,\mpl^4\,m_X^4\,\pi\,s^6}\Biggl[192m_V^8 \left(s-4 m_X^2\right)^4\nonumber\\
&-32m_V^6 s \left(s-4 m_X^2\right)^2 \left(96 m_X^4-68 m_X^2 s+s^2\right)\nonumber\\
&+8m_V^4 s^2 \left(2304 m_X^8-2624 m_X^6 s+1304 m_X^4 s^2-184 m_X^2 s^3+19 s^4\right)\nonumber\\&
-4m_V^2 s^3 \left(768 m_X^8-608 m_X^6 s+368 m_X^4 s^2-148 m_X^2 s^3+13 s^4\right)\nonumber\\&
+s^4 \left(192 m_X^8-32 m_X^6 s+152 m_X^4 s^2-52 m_X^2 s^3+7 s^4\right)\Biggr]
\\
\sigma\left(s\right)_{HH\to XX} =& \frac{\xi^2\,(s-4\,m_X^2)}{480\,\mpl^4\,m_X^4\,\pi\,s^6}\Biggl[192 m_h^8 \left(s-4 m_X^2\right)^4\nonumber\\
&+128 m_h^6 s \left(s-4 m_X^2\right)^2 \left(-24 m_X^4+12 m_X^2 s+s^2\right)\nonumber\\&
+32 m_h^4 s^2 \left(576 m_X^8-496 m_X^6 s+176 m_X^4 s^2-31 m_X^2 s^3+6 s^4\right)\nonumber\\
&+8 m_h^2 s^3 \left(-384 m_X^8+224 m_X^6 s-4 m_X^4 s^2-31 m_X^2 s^3+6 s^4\right)\nonumber\\
&+s^4 \left(192 m_X^8-32 m_X^6 s+152 m_X^4 s^2-52 m_X^2 s^3+7 s^4\right)\Biggr].
\end{align}

In the limit when all SM particles are massless, we find

\begin{equation}
 \sigma(s)=
 \frac{\xi^2}{\pi\,\mpl^4}\left(\frac{13 s^3}{540 m_X^4}+\frac{115 s^2}{432 m_X^2}+\frac{5s}{3 }+\frac{344 m_X^6}{45 s^2}-\frac{194 m_X^4}{27  s}-\frac{71 m_X^2}{27 }\right)
\simeq 0.02\frac{\xi^2}{\pi\,\mpl^4}\,\frac{s^3}{m_X^4}\,,
\end{equation}

\noindent assuming $s\gg m_X^2$.


\bibliographystyle{JHEP}
\bibliography{Bibliography}

\end{document}